\documentclass[11pt]{article}

\usepackage{latexsym,amsfonts,amsmath,amssymb,enumerate,epsfig}
\usepackage{theorem}
\usepackage{subfig}
\usepackage{mathrsfs}
\usepackage{ulem}
\usepackage{tikz}
\usepackage[all]{xy}

\usepackage{color}

\numberwithin{equation}{section}

\newcommand{\mb}[1]{\quad\mbox{#1}\quad}
\newcommand{\beq}{\begin{equation}}
\newcommand{\eeq}{\end{equation}}
\newcommand{\beu}{\begin{equation*}}
\newcommand{\eeu}{\end{equation*}}
\newcommand{\bea}{\begin{eqnarray}}
\newcommand{\eea}{\end{eqnarray}}
\newcommand{\beano}{\begin{eqnarray*}}
\newcommand{\eeano}{\end{eqnarray*}}
\newcommand{\bmx}{\begin{pmatrix}}
\newcommand{\emx}{\end{pmatrix}}

\newcommand{\nonu}{\nonumber\\}

\newcommand{\llangle}{\langle\!\langle}
\newcommand{\rrangle}{\rangle\!\rangle}
\newcommand{\steady}{|{\cal S}\rangle} 
\begin{document}
 
\begin{center}

 {\LARGE  {\sffamily Integrable dissipative exclusion process:\\[1ex]
  Correlation functions and physical properties.} }\\[1cm]

\vspace{10mm}
  
{\Large 
 N. Crampe$^{a,}$\footnote{nicolas.crampe@univ-montp2.fr}, 
 E. Ragoucy$^{b,}$\footnote{eric.ragoucy@lapth.cnrs.fr},
 V. Rittenberg$^{c,}$\footnote{vladimirrittenberg@yahoo.com}
 and M. Vanicat$^{b,}$\footnote{matthieu.vanicat@lapth.cnrs.fr}}\\[.41cm] 
{\large $^a$ Laboratoire Charles Coulomb (L2C), UMR 5221 CNRS-Universit\'e de Montpellier,\\[.242cm]
Montpellier, F-France.}
\\[.42cm]
{\large $^{b}$ Laboratoire de Physique Th{\'e}orique LAPTh,
 CNRS and Universit{\'e} de Savoie,\\[.242cm]
   9 chemin de Bellevue, BP 110, F-74941  Annecy-le-Vieux Cedex, 
France. }
\\[.42cm]
{\large $^{c}$ Physikalisches Institut der Universitaet Bonn,\\[.242cm]
Nussallee 12
D-53115 Bonn,
Federal Republic of Germany.}
\end{center}
\vfill

 \begin{abstract}
We study a one-parameter generalization of the symmetric simple exclusion process on a one-dimensional lattice.
In addition to the usual dynamics (where particles can hop with equal rates to the left or to the right with an exclusion constraint),
annihilation and creation of pairs can occur. The system is driven out of equilibrium by two reservoirs at the boundaries. 

In this
setting the model is still integrable: it is related to the open XXZ spin chain through a gauge transformation. This allows us to compute 
the full spectrum of the Markov matrix using Bethe equations. 

We also show that the stationary state can be expressed 
in a matrix product form permitting to compute the multi-points correlation functions as well as the mean value of the lattice  and  the 
creation-annihilation currents. 

Finally the variance of the lattice current is computed for a finite size system. 
In the thermodynamic limit, it matches  
 the value obtained from the associated macroscopic fluctuation theory. 
\end{abstract}

\vfill\vfill
\rightline{LAPTh-014/16}
\rightline{March 2016}

\newpage
\pagestyle{plain}
\section*{Introduction}

In the context of  biology (going from the microscopic size describing molecular dynamics in the cell 
to the macroscopic one for the evolution of different populations in competition),  chemistry,  physics and
 mathematics, the reaction-diffusion models have been intensively studied. To understand the nature of the stationary
state of such models, driven lattice gas models have been proposed \cite{KLS,Zia,Schutz2,PaulK} and then their generalisation
with dissipation \cite{EF,Racz,HRS,andrea,DFL,B,LL,SL,F,FP}.
For particular choices of the parameters, one-dimensional diffusive gas with dissipation 
can be mapped equivalently to a free fermions problem which can be solved easily 
\cite{BGS,doering,GNS,GS,schutz,oliveira,SN,MB,AM,Alcaraz1,HKP,HSP,SSS}. 

Recently \cite{CRV} we have introduced a new one-dimensional integrable stochastic model. Most of the known integrable stochastic models are derived starting with representations of
quotients of the Hecke \cite{AR2} or  Brauer \cite{brauer} algebras. The new integrable
stochastic model is of a different kind. It was pointed out to us by Pyatov \cite{pyatov} that in fact one deals with a special representation of a quotient of the Birman-Murakami-Wentzel algebra. This is new and it is still an open question how to generalize this model.
It is the aim of this paper to understand what are the physical properties of the model having in mind possible generalizations. This task is simplified by the observation \cite{CRV} that the probability distribution function in the stationary state can be written in terms of the matrix product Ansatz. The dynamics of the model can be obtained using as usual, the Bethe Ansatz or, as shown in the paper, the Macroscopic Fluctuation Theory (MFT) \cite{BL,BL2}.

The model is a one-parameter deformation of Symmetric Simple Exclusion Process (SSEP) model, allowing pairs of particles to be generated or get annihilated with equal rates. These rates are fixed by the parameter. We call the model DiSSEP where Di stands for dissipative. Similarly to SSEP one can add sources and sinks at the end of the system keeping the integrability of the model.
What can we expect to be the physics of the model? Since hopping takes place in a symmetric way, in the thermodynamic limit one can have only weakly correlation functions (the correlators see only the boundaries and not the distance between the particles). Finite-size effects, however, can be new since one can make the parameter dependent on the size of the system.

The stationary state is expressed using a matrix product ansatz, which allows us to compute the correlation functions and the mean value of the currents.
A new feature is that we have two currents: one of them given by
particles crossing the bonds and a second one given by particles leaving the
system.
A new feature appears also here in comparison to the SSEP with sinks and sources: the values of these currents are not homogeneous and depend on the site where 
we measure them. We describe in details the different behaviors of these physical quantities depending on the 
boundary parameters in the thermodynamic limit. 

Using recursive relations between the correlation functions, we succeed in giving a closed analytical relation for the variance of
the lattice current. This variance depends also on the site of the lattice which makes it more involved in comparison to the SSEP. 
A by-product of our research was the check of the applicability of the MFT to stochastic processes with creation and annihilation of particles. We have have shown that in a particular case when the dynamics is of diffusive kind the results obtained from lattice calculations and MFT coincide.
This result is relevant since up to now the validity of MFT developed in \cite{Bertini1,Bertini2,GianniRevue} was confirmed only in purely diffusive models such as SSEP \cite{DCairns}.

Finally, using Bethe ansatz approach, the spectrum of the associated Markow matrix is completely characterized by Bethe equations.
Comparing with the exact diagonalisation of the Markov matrix for small lattices, we can determine the Bethe state 
allowing us to compute the greatest non vanishing eigenvalue. In this way, we can compute the spectral gap
for large lattices (up to 150 sites) and we conjecture its thermodynamic limit by extrapolation. 
\\

The plan of the paper is as follows. In section \ref{sec:des}, we describe precisely the DiSSEP, 
its symmetries and its associated Markov matrix. Then, in section \ref{sect:lambda1}, we remark that for a particular choice of this parameter
the eigenvectors and the eigenvalues become simpler and can be all computed explicitly. This allows us to infer the gap and the large deviation function for 
the current entering in the system from one of the reservoirs. Then, we start the study of the general case. In section \ref{section:stationary_state},
we present the computation of the stationary state using the matrix product ansatz method that provides us analytical expressions for correlation functions.
We compute also exactly the variance of the lattice current (that depends on the site where it is measured). We take the thermodynamic limit of the model in  section \ref{sect:thermo}:
we show that the additional parameter must be rescaled with the length in order to have a competition between the hopping and the evaporation.
We deduce from the previous microscopic computations the exact expressions for the densities, the currents and the variance of the lattice current.
In section \ref{sect:MFT}, we show that the latter is in agreement with the results obtained from macroscopic fluctuation theory. Finally, we relate in section \ref{sec:bethe} the Markov matrix of the DiSSEP to the Hamiltonian of the XXZ spin chain with triangular boundaries \cite{Bel}. 
The eigenvalues of this Hamiltonian have been computed previously using Bethe ansatz methods. We present the associated Bethe equations and 
we solve them to get the spectral gap. In Section \ref{conclu} we summarize our results and give an outlook for further research.

\section{A one parameter deformation of the SSEP \label{sec:des}}

\subsection{Description of the model}

\begin{figure}[htb]
\begin{center}
 \begin{tikzpicture}[scale=0.7]
\draw (-2,0) -- (12,0) ;
\foreach \i in {-2,-1,...,12}
{\draw (\i,0) -- (\i,0.4) ;}
\draw[->,thick] (-2.4,0.9) arc (180:0:0.4) ; \node at (-2.,1.8) [] {$\alpha$};
\draw[->,thick] (-1.6,-0.1) arc (0:-180:0.4) ; \node at (-2.,-0.8) [] {$\gamma$};
\draw  (1.5,0.5) circle (0.3) [fill,circle] {};
\draw  (4.5,0.5) circle (0.3) [fill,circle] {};
\draw  (5.5,0.5) circle (0.3) [fill,circle] {};
\draw  (8.5,3.1) circle (0.3) [fill,circle] {};
\draw  (9.5,3.1) circle (0.3) [fill,circle] {};
\draw[->,thick] (1.4,1) arc (0:180:0.4); \node at (1.,1.8) [] {$1$};
\draw[->,thick] (1.6,1) arc (180:0:0.4); \node at (2.,1.8) [] {$1$};
\node at (5,1.1) [rotate=-90] {$\Big{\{}$};
\draw[->,thick] (5,1.3) -- (5,2.8); \node at (5.4,2.2) [] {$\lambda^2$};
\node at (9,2.5) [rotate=90] {$\Big{\{}$};
\draw[->,thick] (9,2.3) -- (9,0.8); \node at (9.4,1.6) [] {$\lambda^2$};
\draw[->,thick] (11.6,1) arc (180:0:0.4) ; \node at (12.,1.8) [] {$\beta$};
\draw[->,thick] (12.4,-0.1) arc (0:-180:0.4) ; \node at (12.,-0.8) [] {$\delta$};
 \end{tikzpicture}
 \end{center}
 \caption{Dynamical rules of the model.}
 \label{fig:rdm}
\end{figure}
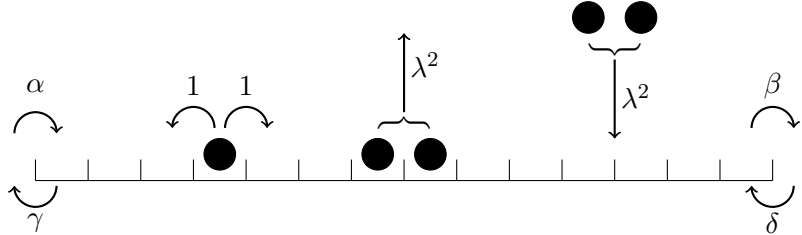

 
 We present here the stochastic process, DiSSEP.
 It describes particles
 evolving on a one-dimensional lattice composed of $L$ sites and connected with two reservoirs at different densities on its extremities.
 There is a Fermi-like exclusion principle: there is at most one particle per site. Hence a configuration $\mathcal{C}$ of the system can be formally
 denoted by an $L$-tuple $(n_1,\dots,n_L)$ where $n_i=0$ if there is no particle at site $i$ and $n_i=1$ if the site is occupied. 
 During each infinitesimal time $dt$, a particle in the bulk can jump to the left or to the right neighboring site with probability $dt$ if it is unoccupied.
 A pair of neighbor particles can also be annihilated with probability $\lambda^2 \times dt$ and be created on unoccupied neighbor sites with probability
 $\lambda^2 \times dt$ (see figure \ref{fig:rdm}). Let us mention that there is a slight change of notation in comparison to \cite{CRV} in order to make  
 the limit to the SSEP easier. At the two extremities of the lattice the dynamics is modified to take in account 
 the interaction with the reservoirs: at the first site (connected with the left reservoir), during time $dt$, a particle is injected with probability $\alpha \times dt$
 if the site is empty and extracted with probability $\gamma \times dt$ if it is occupied. The dynamics is similar at last site (connected
 with the right reservoir) with injection rate $\delta$ and extraction rate $\beta$.
 The dynamical rules can be summarized in the following table where $0$ stands for vacancy and $1$ stands for a particle.
 The transition rates between the configurations are written above the arrows.
 \begin{equation} \label{eq:rules}
 \begin{array}{|c |c| c| }
 \hline \text{Left} & \text{Bulk} & \text{Right} \\
 \hline
 0\, \xrightarrow{\ \alpha \ }\, 1&  01\, \overset{\ 1\ }{\longleftrightarrow} \,10&1\, \xrightarrow{\ \beta \ } \,0\\
 1\, \xrightarrow{\ \gamma \ }\, 0&00\, \overset{\ \lambda^2 \ }{\longleftrightarrow}\, 11&0\, \xrightarrow{\ \delta\ }\, 1\\ \hline
 \end{array}
 \end{equation}
We choose the coefficient  of condensation and evaporation to be $\lambda^2$ and not $\lambda$ for later convenience. 
Let us remark that the SSEP is recovered when the creation/annihilation rate $\lambda^2$ vanishes. 
The limit $\lambda^2\rightarrow\infty$ provides a model with only condensation and evaporation.

The system is driven out of equilibrium 
by the boundaries. As shown in section \ref{section:stationary_state}, there are particle currents in the stationary state for generic 
boundary rates $\alpha$, $\beta$, $\gamma$ and $\delta$. We will see below \eqref{1pt-corr-limit} that these choices of rates describe particle
reservoirs with densities
\begin{equation} \label{densities}
 \rho_a=\frac{\alpha}{\alpha+\gamma}, \quad \text{and} \quad \rho_b=\frac{\delta}{\beta+\delta}.
\end{equation}

\hfill\break 
 \textsl{Remark:} The system will reach a thermodynamic equilibrium if and only if both the densities of the reservoirs are equal to $1/2$,
  that is $\alpha=\gamma$ and $\beta=\delta$.
 Indeed, the detailed balance is only satisfied in this case.

\null

\subsubsection*{Symmetries of the model}
We can make the following observations:
\begin{itemize}
\item Since we chose the evaporation/condensation rate to be $\lambda^2$, all the results shall be 
invariant under the transformation $\lambda\rightarrow -\lambda$. 
\item The left/right symmetry of the chain is given by the transformations $\alpha\leftrightarrow\delta$, $\gamma\leftrightarrow\beta$ and a change of numbering of the sites $i\to L+1-i$.
\item The vacancy-particle symmetry translates into $\alpha\leftrightarrow\gamma$ and $\delta\leftrightarrow\beta$.
\end{itemize}

\subsection{Markov matrix}

We denote by $P_t(\mathcal{C})$ the probability for the system to be in configuration $\mathcal{C}$ at time $t$. $P_t(\mathcal{C})$ obeys
the master equation
\begin{equation}\label{eq:ME}
\frac{d P_t(\mathcal{C})}{dt}=\sum_{\mathcal{C}'\neq \mathcal{C}}M(\mathcal{C},\mathcal{C}')P_t(\mathcal{C}')
-\sum_{\mathcal{C}'\neq \mathcal{C}}M(\mathcal{C}',\mathcal{C}) P_t(\mathcal{C})
 = \sum_{\mathcal{C}'}M(\mathcal{C},\mathcal{C}')P_t(\mathcal{C}') \,,
\end{equation}
where $M(\mathcal{C},\mathcal{C}')$ is the transition rate from the configuration $\mathcal{C}'$ to the configuration $\mathcal{C}$. The second
equality is obtained by setting $M(\mathcal{C},\mathcal{C})=-\sum_{\mathcal{C}'\neq \mathcal{C}}M(\mathcal{C}',\mathcal{C})$.
This equation can be recast in a compact way: let us set 
\begin{equation}
 |P_t\rangle=\left(
 \begin{array}{c}
  P_t(\ (0,\dots,0,0,0)\ )\\
   P_t(\ (0,\dots,0,0,1)\ )\\
   P_t(\ (0,\dots,0,1,0)\ )\\
   \vdots\\
   P_t(\ (1,\dots,1,1,1)\ )
 \end{array}
\right)
=\sum_{n_1,\dots,n_L\in \{0,1\}}  P_t(\ (n_1,\dots,n_{L})\ )~ |n_1\dots n_L\rangle
\end{equation}
with $|n_1\dots n_L\rangle=|n_1\rangle\otimes\dots\otimes|n_L\rangle$,
$|0\rangle=\left(
 \begin{array}{c}
 1\\0
 \end{array}
\right)$ and $|1\rangle=\left(
 \begin{array}{c}
 0\\1
 \end{array}
 \right)$.
 In this formalism, each site of the lattice corresponds to one copy of $\mathbb{C}^2$ in the tensorial space
 $\left(\mathbb{C}^2\right)^{\otimes L}$.
 The master equation \eqref{eq:ME} is then rewritten as
\begin{equation}\label{eq:MEv}
 \frac{d|P_t\rangle}{dt}=M\ |P_t\rangle.
\end{equation}

The Markov matrix $M$ corresponding to the dynamical rules \eqref{eq:rules} can be expressed as
\begin{equation} \label{Markov-matrix}
 M=B_1+\sum_{k=1}^{L-1}w_{k,k+1}+\overline{B}_L,
\end{equation}
where the local jump operators $B$, $\overline{B}$ and $w$ are given by
\begin{equation} \label{eq:localw8vertex}
B =\left( \begin {array}{cc} 
-\alpha&\gamma\\ 
\alpha&-\gamma
\end {array} \right)\quad ,\qquad  w=\left( \begin {array}{cccc} 
-\lambda^2&0&0&\lambda^2\\ 
0&-1&1&0\\
0&1&-1&0\\
\lambda^2&0&0&-\lambda^2
\end {array} \right)
\quad\text{,}\qquad
\overline{B} =\left( \begin {array}{cc} 
-\delta&\beta\\ 
\delta&-\beta
\end {array} \right)\;.
\end{equation}
The subscripts in \eqref{eq:MEv} indicate on which sites the local operators are acting.

\section{Study of the particular case $\lambda=1$\label{sect:lambda1}}

Before studying the general model, we focus on the case $\lambda=1$, 
where the calculations simplify drastically: it corresponds to the free fermion point of the model we introduced.

\subsection{Eigenvectors and relaxation rate}
For $\lambda=1$, all the eigenvalues and the eigenvectors can be computed easily. Indeed, for $L\geq3$, the $2^L$ eigenvectors are characterized by the 
set $\boldsymbol{\epsilon}=(\epsilon_1,\epsilon_2,\dots,\epsilon_L)$ with $\epsilon_i=\pm 1$ and are given by 
\begin{equation}
 \Omega(\boldsymbol{\epsilon})=v(\epsilon_1,\epsilon_2,\alpha,\gamma)\otimes\begin{pmatrix}
                                                   1\\ \epsilon_2
                                                  \end{pmatrix}\otimes\begin{pmatrix}
                                                   1\\ \epsilon_3
                                                  \end{pmatrix}\otimes\cdots\otimes\begin{pmatrix}
                                                   1\\ \epsilon_{L-1}
                                                  \end{pmatrix}\otimes v(\epsilon_L,\epsilon_{L-1},\delta,\beta)
\end{equation}
where $v(\epsilon,\epsilon',\mu,\nu)=\begin{pmatrix}  \epsilon'+\nu\\  1+\mu+f(\epsilon,\epsilon',\mu+\nu) \end{pmatrix}$
and $f(\epsilon,\epsilon',\tau)=\epsilon\epsilon'-1-\frac{\tau}{2}(1-\epsilon)$. The corresponding eigenvalues are
\begin{equation}
 \Lambda(\boldsymbol{\epsilon})=f(\epsilon_1,\epsilon_2,\alpha+\gamma)
 + \sum_{j=2}^{L-2}( \epsilon_j\epsilon_{j+1}-1)+f(\epsilon_L,\epsilon_{L-1},\delta+\beta)\;.
\end{equation}
Let us remark that the ASEP on a ring with Langmuir kinetics has been treated similarly in \cite{sato}. 
From the previous results, we deduce that the stationary state is 
\begin{equation}
 \Omega(+,+,\dots,+)= \frac{1}{Z_L}\ \   \begin{pmatrix}  1+\gamma\\ 1+\alpha
                                                  \end{pmatrix}\otimes\begin{pmatrix}
                                                   1\\ 1
                                                  \end{pmatrix}^{\otimes L-2}\otimes\begin{pmatrix}
                                                   1+\beta\\ 1+\delta
                                                  \end{pmatrix} \end{equation}
where $Z_L=2^{L-2}(2+\alpha+\gamma)(2+\beta+\delta)$ is the normalisation such that the entries be probabilities.

From this stationary state, we can compute the mean value of the injected current by the left reservoir (resp. by the right reservoir)
\begin{equation}\label{eq:jmean}
 j_{\text{left}}=\frac{\alpha-\gamma}{2+\alpha+\gamma} \qquad(\text{resp. }  j_{\text{right}}=\frac{\delta-\beta}{2+\delta+\beta})\;.
\end{equation}
We see that the current 
has the sign of $\alpha-\gamma$ (resp. $\delta-\beta$). As expected, it goes to the left when extraction is promoted, and to the right when injection is proeminent. It vanishes for $\alpha=\gamma$. The lattice current in the bulk vanishes.

We can also compute easily the first excited state whose eigenvalue provides the relaxation rate. Indeed, the greatest non vanishing eigenvalue is
\begin{equation}
 \begin{cases}
  G=-4 & \text{if } \alpha+\gamma>2 \mb{and} \beta+\delta>2\\
 G=-2-\beta-\delta& \text{if } \alpha+\gamma>2\mb{and} \beta+\delta<2\\
 G=-2-\alpha-\gamma& \text{if } \alpha+\gamma<2 \mb{and} \beta+\delta>2\\
 G=-\alpha-\gamma-\beta-\delta& \text{if } \alpha+\gamma<2\mb{and}\beta+\delta<2
 \end{cases}
\end{equation}
These results shall be generalized in section \ref{sec:gap} for any $\lambda$ using the Bethe equations. 
The general result displayed on Figure \ref{fig:g3} matches the above values of the gap for $\lambda=1$ ($\phi=0$).

\subsection{Current large deviation function}

For this particular choice of $\lambda$, it is also possible to get the generating function of the cumulants of the current 
entering in the system from the left reservoir (the same result is also obtained by symmetry for the right reservoir). 
For general $\lambda$, one obtains the variance in section \ref{sec:flu}. It is well established that this generating function is the greatest eigenvalue 
of the following deformed Markov matrix \cite{DonVar}
\begin{equation} \label{Markov-matrix_s}
 M=B_1(s)+\sum_{k=1}^{L-1}w_{k,k+1}+\overline{B}_L,
\end{equation}
where the local jump operator $B(s)$ is deformed as follows
\begin{equation} \label{eq:localBs}
B =\left( \begin {array}{cc} 
-\alpha&\gamma e^{-s}\\ 
\alpha e^s&-\gamma
\end {array} \right)\;.
\end{equation}
One can show that the greatest eigenvalue is given by
\begin{equation}\label{eq:ES}
 E(s)=-\frac{2+\alpha+\gamma}{2}+\frac{1}{2}\sqrt{4+4\alpha e^s+4\gamma e^{-s}+(\alpha+\gamma)^2}
\end{equation}
with the eigenvector 
\begin{equation}
 \Omega(s)=     \begin{pmatrix}  1+\gamma e^{-s}\\ 1+\alpha +E(s)
                                                  \end{pmatrix}\otimes\begin{pmatrix}
                                                   1\\ 1
                                                  \end{pmatrix}^{\otimes L-2}\otimes\begin{pmatrix}
                                                   1+\beta\\ 1+\delta
                                                  \end{pmatrix}\;. \end{equation}

\begin{figure}[htbp]
\begin{center}
\includegraphics[width=80mm,height=80mm]{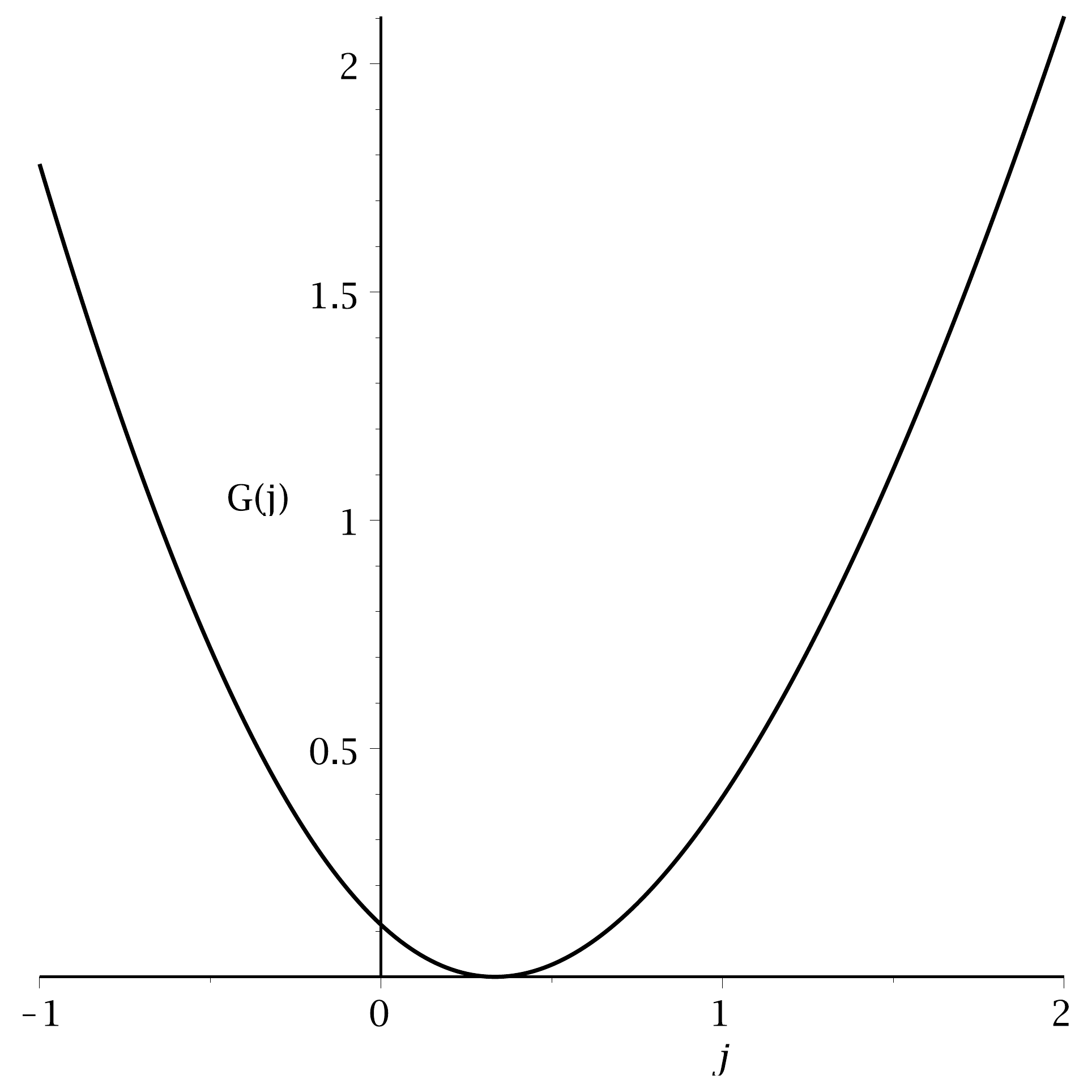}
\end{center}
\caption{Example of large deviation function $G(j)$ (on the plot $\alpha=2,\gamma=0.5$). \label{fig:large_devia}}
\end{figure}

The rate-function $G(j)$ associated to the current is the Legendre transformation of this generating function of the cumulants:
 \begin{equation} 
G(j)=s^* j-E(s^*)\mb{,}\frac{d}{ds}E(s)\Big|_{s=s^*}=j 
\end{equation}
Then its explicit form is given by
\begin{equation}\label{full-G}
 G(j)=1+\frac{\alpha+\gamma}{2}-\sqrt{1+\Delta(j)+\left(\frac{\alpha+\gamma}{2}\right)^2}
 +j\ln\left(\frac{\Delta(j)}{2\alpha}+\frac{j}{\alpha}\sqrt{1+\Delta(j)+\left(\frac{\alpha+\gamma}{2}\right)^2} \right)
\end{equation}
where 
\begin{equation}
 \Delta(j)=2j^2+\sqrt{4(\alpha\gamma+j^2+j^4)+j^2(\alpha+\gamma)^2}\;.
\end{equation}
Let us stress that \eqref{full-G} represents an exact result on the large deviation function of the current on the left boundary.
The function $G(j)$ is convex and vanishes when $j$ is equal to the mean value of the current on the left boundary given by \eqref{eq:jmean} as expected, 
see figure \ref{fig:large_devia}. Note that it is not Gaussian.

\section{Stationary state and physical observables \label{section:stationary_state}}
We now turn to the general case (i.e. for generic $\lambda$). Although one cannot perform all the calculation presented in  the $\lambda=1$ case, 
one can still obtain interesting analytical results, in particular when focusing on the stationary state.

\subsection{Matrix product anstatz}
The integrability of the model reflects in the simple structure of its stationary state which can be built in a matrix product form.
The power of this technique was revealed in a pioneering work \cite{DEHP}, where the phase diagram of the totally asymmetric simple
exclusion process (TASEP) was computed analytically, using a matrix product expression of the steady state wave function.  
It has led to numerous works and generalizations, among them can be mentioned the multi-species TASEP \cite{PEM,EFM,CMRV}
and more complicated reaction-diffusion processes \cite{HSP,Isaev}. A review of these results can be found in \cite{MartinRev}.
In the framework of integrable Markov matrix, the stationary state can be expressed through a matrix product ansatz which take a simple form: the complete 
algebraic structure can be determined thanks to the Zamolodchikov-Faddeev and Ghoshal-Zamolodchikov relation as it was first seen in 
\cite{Sasamoto2}. A systematic construction of the Matrix product ansatz for integrable systems can be found in \cite{CRV}. 

It was shown in \cite{CRV} that 
the steady state $\steady$ of the present model can be built as follows
\begin{equation} \label{steady-state}
 \steady = \frac{1}{Z_L}\llangle W| \left( \begin{array}{c}
                                            E \\
                                            D
                                           \end{array} \right)^{\otimes L} |V\rrangle,
\end{equation}
where $Z_L$ is the normalization factor
\begin{equation}
 Z_L=\llangle W|(E+D)^L|V\rrangle.
\end{equation}
The algebraic elements $E$ and $D$ belongs to an algebra composed of three generators $E$, $D$ and $H$. The commutation relations 
between these generators are given by 
\begin{equation} \label{eq:comEDH}
 [D,E]=EH+HD, \quad \text{and} \quad [H,E]=[H,D]=\lambda^2(D^2-E^2).
\end{equation}
These relations are equivalent to the very useful telescopic relation
\begin{equation} \label{telescopic-bulk}
w \left( \begin{array}{c}
         E \\
         D
        \end{array} \right) \otimes
        \left( \begin{array}{c}
         E \\
         D
               \end{array} \right) = 
        \left( \begin{array}{c}
         E \\
         D
               \end{array} \right) \otimes 
        \left( \begin{array}{c}
         -H \\
         H
        \end{array} \right)-
        \left( \begin{array}{c}
         -H \\
         H
               \end{array} \right) \otimes 
        \left( \begin{array}{c}
         E \\
         D
        \end{array} \right)\;.
\end{equation}
Notice here that, in contrast with the SSEP case (see subsection \ref{comparison-SSEP}) where $H$ is a scalar, the commutation relations
between $H$ and $E, \ D$ are not trivial.
The action of the generators $E$, $D$ and $H$ on the boundary vectors $\llangle W|$ and $|V\rrangle$ is given by
\begin{equation}
 \llangle W|\left(\alpha E-\gamma D \right)=\llangle W|H, \quad \text{and} \quad \left(\delta E-\beta D\right)|V\rrangle = -H|V\rrangle.
\end{equation}
It is equivalent to
\begin{equation} \label{telescopic-boundaries}
 \llangle W|B \left( \begin{array}{c}
         E \\
         D
               \end{array} \right) = 
 \llangle W|\left( \begin{array}{c}
         -H \\
         H
        \end{array} \right), \quad \text{and} \quad 
\overline{B} \left( \begin{array}{c}
              E \\
              D
                    \end{array} \right)|V\rrangle = 
 -\left( \begin{array}{c}
         -H \\
         H
        \end{array} \right)|V\rrangle.       
\end{equation}
It is straightforward to see using \eqref{telescopic-bulk} and \eqref{telescopic-boundaries} that 
$M\steady=0$ with $M$ defined by \eqref{Markov-matrix} and $\steady$ defined by \eqref{steady-state}. Indeed, we get a telescopic sum.
We showed also in \cite{CRV} that the algebra is consistent with the boundary equations by giving an explicit representation of the generators 
$E$, $D$, $H$ and of the boundary vectors $|V\rrangle$
and $\llangle W|$.
We have tried, without success, to get the probability distribution function for the stationary state in the case of the asymmetric hopping rates. This would have been very interesting. 
It is surprising since our matrix product ansatz is close to the one used to solve the symmetric simple exclusion process. Our failure is probably due to the non-integrability of the system in this case.

Assuming that $\lambda$ is non vanishing,  we make a change of basis from the generators $E$, $D$ and $H$ to $G_1$, $G_2$ and $G_3$ as follows
\begin{equation} \label{change-basis}
 \left\{ \begin{aligned}
          & E=G_1+G_2+G_3, \\
          & D=G_2-G_1-G_3, \\
          & H=2\lambda(G_3-G_1).
         \end{aligned}
 \right.
\end{equation}
This change of basis will simplify the calculation, moreover it is natural in the general framework developed in \cite{CRV}.
The commutation relations between $E$, $D$ and $H$ \eqref{eq:comEDH} are equivalent to
\begin{equation} \label{eq:comG}
 [G_1,G_3]=0, \quad G_2G_1=\phi \ G_1G_2, \quad \text{and} \quad G_3G_2=\phi \ G_2G_3, \quad  \text{with} \quad \phi=\frac{1-\lambda}{1+\lambda}.
\end{equation}
The relations on the boundaries become
\begin{equation} \label{eq:bordsG}
 \left\{ \begin{aligned}
          & \llangle W |\big( G_1-c\;G_2-a\;G_3 \big)=0\,, \\
          &\big( G_3-b\;G_1-d\;G_2 \big) | V \rrangle =0
         \end{aligned}
 \right.
 \mb{with}
 \left\{ \begin{aligned}
& a=\frac{2\lambda-\alpha-\gamma}{2\lambda+\alpha+\gamma}\,,\quad 
& c=\frac{\gamma-\alpha}{2\lambda+\alpha+\gamma}\,,
\\
& b=\frac{2\lambda-\delta-\beta}{2\lambda+\delta+\beta}\,, 
& d=\frac{\beta-\delta}{2\lambda+\delta+\beta}\,.
 \end{aligned}
 \right.
\end{equation}

We can now give the main result of this paper. Indeed, in this new basis, it is possible to compute 
a closed expression for any word, for $p,q,r=0,1,2,\dots$,
\begin{equation}\label{eq:G123}
 \frac{\llangle W|G_1^pG_2^qG_3^r|V\rrangle}{\llangle W|G_2^{p+q+r}|V\rrangle}=
 \frac{\displaystyle \prod_{\ell=0}^{p-1}(c\ \phi^{p-1-\ell}+ad\ \phi^{q+r+\ell})\ \prod_{n=0}^{r-1}(d\ \phi^{r-1-n}+bc\ \phi^{q+p+n})}
{\displaystyle\prod_{k=q}^{p+q+r-1}(1-ab\ \phi^{2k})}\ ,
\end{equation}
where by convention $\displaystyle \prod_{n=0}^{-1} (\cdot)=1$. {Let us mention that this formula is not valid if $ab\phi^k=1$ which 
occurs for $\lambda=0$ (SSEP case) or $\lambda\rightarrow \infty $. However these limits can be performed  for the physical quantities (see section \ref{comparison-SSEP}). }

\paragraph{Proof of the formula \textbf{\eqref{eq:G123}}.}

In order to compute $\llangle W|G_1^pG_2^qG_3^r|V\rrangle$, we use a change of generators defined as follows
\begin{equation} \label{eq:GtoLR}
 L_i=\frac{G_1}{\phi^i}-aG_3\phi^i, \quad \text{and} \quad R_k=\frac{G_3}{\phi^k}-bG_1\phi^k.
\end{equation}
This is built so that $L_i$ and $R_k$ fulfill the following relations (derived straightforwardly from \eqref{eq:comG} and \eqref{eq:bordsG})
\begin{eqnarray} \label{eq:relLR}
&&G_2^iL_i=L_0G_2^i, \quad R_kG_2^k=G_2^kR_0, \quad [L_i,G_1]=[L_i,G_3]=[R_k,G_1]=[R_k,G_3]=0\qquad\\
&\text{and}&  \llangle W|L_0=c\llangle W|G_2, \quad R_0|V\rrangle = dG_2|V\rrangle.\label{eq:relLR2}
\end{eqnarray}
The change of generators \eqref{eq:GtoLR} can be inverted to get
\begin{equation} \label{eq:LRtoG}
 G_1=\frac{\phi^{k+i}}{1-ab\phi^{2(k+i)}}\left(\frac{L_i}{\phi^k}+a\phi^iR_k\right), \quad \text{and} \quad 
 G_3=\frac{\phi^{k+i}}{1-ab\phi^{2(k+i)}}\left(b\phi^kL_i+\frac{R_k}{\phi^i}\right).
\end{equation}
We can now begin the computation
\begin{eqnarray*}
 \llangle W|G_1^pG_2^qG_3^r|V\rrangle & = & \frac{\phi^q}{1-ab\phi^{2q}}\llangle W|\left(\frac{L_0}{\phi^q}+aR_q\right)G_1^{p-1}G_2^qG_3^r|V\rrangle \\
 & = &  \frac{1}{1-ab\phi^{2q}}\llangle W|cG_2G_1^{p-1}G_2^qG_3^r|V\rrangle+\frac{1}{1-ab\phi^{2q}}\llangle W|G_1^{p-1}G_2^qG_3^rad\phi^qG_2|V\rrangle \\
 & = & \frac{c\phi^{p-1}+ad\phi^{q+r}}{1-ab\phi^{2q}}\llangle W|G_1^{p-1}G_2^{q+1}G_3^r|V\rrangle.
\end{eqnarray*}
The first equality is obtained using \eqref{eq:LRtoG} with $i=0$ and $k=q$ to transform the leftmost $G_1$. The second equality relies on the 
relations \eqref{eq:relLR} and \eqref{eq:relLR2}. We get the last one using \eqref{eq:comG}.
This relation is a recursive relation between $\llangle W|G_1^pG_2^qG_3^r|V\rrangle$ and $\llangle W|G_1^{p-1}G_2^{q+1}G_3^r|V\rrangle$ that we 
can iterate to obtain
\begin{equation} \label{eq:recG1}
 \llangle W|G_1^pG_2^qG_3^r|V\rrangle = \left(\prod_{l=0}^{p-1} \frac{c\phi^{p-1-l}+ad\phi^{r+q+l}}{1-ab\phi^{2(q+l)}}\right)\llangle W|G_2^{q+p}G_3^r|V\rrangle.
\end{equation}
Performing similar computations with $G_3$  we obtain the following recursive relation
\begin{eqnarray*}
 \llangle W|G_2^{q+p}G_3^r|V\rrangle & = & \frac{\phi^{q+p}}{1-ab\phi^{2(q+p)}}\llangle W|G_2^{q+p}G_3^{r-1}\left(bL_{q+p}+\frac{R_0}{\phi^{q+p}}\right)|V\rrangle \\
& = & \frac{d\phi^{r-1}+bc\phi^{q+p}}{1-ab\phi^{2(q+p)}}\llangle W|G_2^{q+p+1}G_3^{r-1}|V\rrangle\;,
\end{eqnarray*}
to get
\begin{equation} \label{eq:recG3}
 \llangle W|G_2^{q+p}G_3^r|V\rrangle = \left(\prod_{n=0}^{r-1}\frac{d\phi^{r-1-n}+bc\phi^{q+p+n}}{1-ab\phi^{2(q+p+n)}}\right)\llangle W|G_2^{p+q+r}|V\rrangle.
\end{equation}
Recombining \eqref{eq:recG1} and \eqref{eq:recG3} together, the desired result \eqref{eq:G123} is proved.

Let us stress that since $G_1,G_2,G_3$ form a basis, the knowledge of all words built on them allows us to reconstruct all words built on $E$ and $D$ using the two first relations of \eqref{change-basis}. In particular, we are able to compute exactly physical observables, as it is illustrated below.

\subsection{Calculation of physical observables: correlation functions, currents\label{sec:phys-obs}}
As usual the matrix product ansatz permits to compute the physical observables (see \cite{DEHP}).
For example, the one point correlation function (or density) is expressed as follows in terms 
of the matrix product ansatz
\begin{equation}
 \langle n_i \rangle =  \frac{\llangle W| C^{i-1}DC^{L-i} |V\rrangle}{\llangle W| C^L |V\rrangle}\;,
\end{equation}
where we have used the usual notation $C=E+D$. Let us remark that, in our basis $C=2G_2$.
It is well-known that similar expressions exist also for all the correlations or the currents (see below).

Thanks to expression \eqref{eq:G123} of any words, it is now possible to compute physical observables for DiSSEP.
In particular, we can compute the one point correlation function
\begin{eqnarray} \label{1pt-corr}
 \langle n_i \rangle & = & \frac{1}{2}\frac{\llangle W| G_2^{i-1}(-G_1+G_2-G_3)G_2^{L-i} |V\rrangle}{\llangle W| G_2^L |V\rrangle} \\
& = & \frac{1}{2}-\frac{c\phi^{i-1}+ad\phi^{L+i-2}+d\phi^{L-i}+bc\phi^{2L-i-1}}{2(1-ab\phi^{2L-2})}\:,
\label{1pt-corr-bis}
\end{eqnarray}
the connected two-point correlation function, for $1\leq i<j\leq L$
\begin{eqnarray}\label{2pt-corr}
 \langle n_i n_j \rangle_c & = & \langle n_i n_j \rangle - \langle n_i \rangle \langle n_j \rangle \\
 & = & \frac{\llangle W| C^{i-1}DC^{j-i-1}DC^{L-j} |V\rrangle}{\llangle W| C^L |V\rrangle}
 -\frac{\llangle W| C^{i-1}DC^{L-i} |V\rrangle}{\llangle W| C^L |V\rrangle}\frac{\llangle W| C^{j-1}DC^{L-j} |V\rrangle}{\llangle W| C^L |V\rrangle}\qquad\\
 & = & \frac{\phi^{L+j-i-3}(1-\phi^2)(1+b\phi^{2(L-j)})(1+a\phi^{2(i-1)})(d+bc\phi^{L-1})(c+ad\phi^{L-1})}{4(1-ab\phi^{2(L-1)})^2(1-ab\phi^{2(L-2)})},
 \label{2pt-corr-bis}
\end{eqnarray}
and the connected three-point correlation function, for $1\leq i<j<k\leq L$
\begin{eqnarray} \label{3pt-corr}
 \langle n_i n_j n_k\rangle_c & = & \langle n_i n_j n_k\rangle 
 - \langle n_i \rangle \langle n_j n_k \rangle- \langle n_j \rangle \langle n_i n_k \rangle- \langle n_k \rangle \langle n_i n_j \rangle 
 +2 \langle n_i\rangle\langle n_j \rangle\langle n_k\rangle \\
 &=&-\frac{\phi^{L+k-i-5}(1-\phi^2)^2(1+b\phi^{2(L-k)}) (1+a\phi^{2(i-1)}) (d+bc\phi^{L-1}) (c+ad\phi^{L-1}) }
         {8(1-ab\phi^{2(L-1)})^3(1-ab\phi^{2(L-2)})(1-ab\phi^{2(L-3)})}\nonumber\\
 &\times&\big[ \phi^{L-j}(d+bc\phi^{L-3})(1+2a\phi^{2(j-1)}+ab\phi^{2(L-1)})
 \label{3pt-corr-bis}\\
 &&\quad+\phi^{j-1}(c+ad\phi^{L-3})(1+2b\phi^{2(L-j)}+ab\phi^{2(L-1)})  \big].   \nonumber     
\end{eqnarray}
Remark that for generic $i,j,k$, the two- and three-point correlation functions satisfy both a set of closed linear relations:
\begin{eqnarray} \label{eq:reccurence-corr}
&&(1-\lambda^2) \Big(\langle n_{i-1} n_{j} \rangle_c+\langle n_{i+1} n_{j} \rangle_c+\langle n_{i} n_{j-1} \rangle_c+\langle n_{i} n_{j+1} \rangle_c\Big)
=4(1+\lambda^2) \langle n_{i} n_{j} \rangle_c\,,\\
&&(1-\lambda^2) \Big(\langle n_{i-1} n_{j} n_{k} \rangle_c+\langle n_{i+1} n_{j} n_{k}\rangle_c+\langle n_{i} n_{j-1} n_{k}\rangle_c+\langle n_{i} n_{j+1} n_{k}\rangle_c
+\langle n_{i} n_{j} n_{k-1}\rangle_c\nonumber\\
&&\qquad\qquad+\langle n_{i} n_{j} n_{k+1}\rangle_c\Big)
=6(1+\lambda^2) \langle n_{i} n_{j} n_k\rangle_c
\end{eqnarray}

We can also compute the particle currents. There are two different currents: the lattice current which stands for the number of particles going 
through the bond between site $i$ and $i+1$ per unit of time and the evaporation-condensation current which stands for the number of particle 
evaporating or condensing at sites $i$ and $i+1$ per unit of time. The lattice current is given by
\begin{eqnarray}
 \langle J^{lat}_{i\rightarrow i+1} \rangle & = & \frac{\llangle W|C^{i-1}\left(DE-ED\right)C^{L-i-1}|V\rrangle}{\llangle W|C^L|V\rrangle} \\
 & = & \frac{1-\phi}{2} \frac{bc\phi^{2L-i-2}+d\phi^{L-i-1}-ad\phi^{L+i-2}-c\phi^{i-1}}{1-ab\phi^{2L-2}}.
 \label{mean:lattice}
\end{eqnarray}
Counting positively the pairs of particles which condensate on the lattice and negatively the pairs which evaporate, we get for
the evaporation-condensation current 
\begin{eqnarray}
 \langle J^{cond}_{i, i+1} \rangle & = & \frac{\llangle W|C^{i-1}\lambda^2\left(E^2-D^2\right)C^{L-i-1}|V\rrangle}{\llangle W|C^L|V\rrangle} \\
 & = & \frac{(1-\phi)^2}{2(1+\phi)} \frac{bc\phi^{2L-i-2}+d\phi^{L-i-1}+ad\phi^{L+i-2}+c\phi^{i-1}}{1-ab\phi^{2L-2}}.
 \label{mean:cond}
\end{eqnarray}

Note that the above expressions behave as expected under the three symmetries:
\begin{enumerate}
\item The symmetry $\lambda\to-\lambda$, that translates into $\phi\to1/\phi$, $a\to1/a$, $c\to-c/a$, $b\to1/b$ and $d\to-d/b$, leaves them invariant.
\item The left/right symmetry, that becomes $a\leftrightarrow b$, $c\leftrightarrow d$ and $i\to L+1-i$, changes the sign of the lattice current, keeps the condensation current and the density invariant.
\item The particle-hole symmetry, which reads $a\to a$, $b\to b$, $c\to-c$ and $d\to-d$, changes the sign of both currents and transforms $ \langle n_i \rangle$ into $1- \langle n_i \rangle$.
\end{enumerate}

The physical quantities computed above are not all independent. 
The particle conservation law at site $i$ reads
\begin{equation}\label{eq:part.conservation-discret}
 \langle J^{lat}_{i-1\rightarrow i} \rangle-\langle J^{lat}_{i\rightarrow i+1} \rangle+
 \langle J^{cond}_{i-1, i} \rangle+\langle J^{cond}_{i, i+1} \rangle=0,
\end{equation}
which can be seen on the matrix product ansatz using relations \eqref{eq:comEDH}.
From the identity $[D,E]=[D,C]$ one then deduces 
\beq\label{eq:drho-discret}
\langle J^{lat}_{i\rightarrow i+1} \rangle = \langle n_{i}\rangle - \langle n_{i+1}\rangle ,
\eeq
and using $E^2 -D^2 = C^2-CD-DC$ one  gets
\beq\label{jcond-discret}
\langle J^{cond}_{i, i+1} \rangle=\lambda^2\,\Big(1-\langle n_{i}\rangle - \langle n_{i+1}\rangle\Big).
\eeq
From these three relations, one obtains
\begin{eqnarray} \label{eq:jcond}
&&\langle J^{cond}_{i-1,i} \rangle-\langle J^{cond}_{i, i+1} \rangle 
+\lambda^2 \Big(\langle J^{lat}_{i-1\rightarrow i} \rangle+\langle J^{lat}_{i\rightarrow i+1} \rangle\Big) = 0\\
&&\label{eq:dens}
\langle n_{i-1} \rangle + \langle n_{i+1} \rangle -2\langle n_i \rangle
 +\lambda^2\Big(2-\langle n_{i-1} \rangle - \langle n_{i+1} \rangle -2 \langle n_i \rangle \Big)=0.
\end{eqnarray}

\subsection{Fluctuations of the currents \label{sec:flu}}

As mentioned previously, there are closed linear relations between the two- and three-point correlation functions which allow one to compute
the cumulant of the currents. In this section, we present the computations of the second cumulant of the lattice current between sites $i$ and $i+1$. Let us note that it depends on the site, because of the evaporation-condensation process. As usual for such a purpose \cite{DonVar}, we use the deformed Markovian matrix defined as follows:
\beq
 M^{s}=B_1+\sum_{k=1}^{i-1}w_{k,k+1}+w_{i,i+1}^{s}+\sum_{k=i+1}^{L-1}w_{k,k+1}+\overline{B}_L
 \mb{with}
 w^{s} = \begin{pmatrix} -\lambda^2 & 0 & 0& \lambda^2 \\ 0 & -1 & e^s & 0 \\ 0 & e^{-s} & -1 & 0 \\  \lambda^2 & 0 & 0& -\lambda^2
 \end{pmatrix}.
\end{equation}
Let $ |v^s\rangle$ be the eigenstate of $M^s$ with highest  eigenvalue
\beq
M^s \, |v^s\rangle = \mu(s)\,  |v^s\rangle\,.\label{eigenval-s}
\eeq
$\mu(s)$ is the generating function for the cumulants of the lattice current between sites $i$ and $i+1$.
We introduce {the following notation for vectors}
\bea
\langle\{j\}|&=&\Big(\langle0|+\langle1|\Big)^{\otimes (j-1)} \otimes \langle1|\otimes \Big(\langle0|+\langle1|\Big)^{\otimes (L-j)}\\
\langle\{j,k\}|&=&\Big(\langle0|+\langle1|\Big)^{\otimes (j-1)} \otimes \langle1|\otimes \Big(\langle0|+\langle1|\Big)^{\otimes (k-j-1)} \otimes \langle1|\otimes \Big(\langle0|+\langle1|\Big)^{\otimes (L-k)}\\
\vdots && \nonumber
\eea
In words, $\langle\{j_1,j_2,...,j_M\}|$ represents configurations with one particle at site $j_1$, $j_2$, ..., $j_M$, and anything else  on the other sites. Remark that this definition applies whatever the order on $j_1,...,j_M$, and thus extends the one given in the above equations.  
By extension, we note $\langle\emptyset|=\big(\langle0|+\langle1|\big)^{\otimes L}$. Then, we define the components:
\beq
T_j(s)\equiv T_j = \frac{\langle\{j\} |v^s\rangle}{\langle\emptyset|v^s\rangle}
\mb{;}
U_{jk}(s)\equiv U_{jk} =  \frac{\langle\{j,k\}|v^s\rangle}{\langle\emptyset|v^s\rangle}
\mb{and}
V_{jkl}(s)\equiv V_{jkl} =  \frac{ \langle\{j,k,l\}|v^s\rangle}{\langle\emptyset|v^s\rangle}.
\eeq
{Note that by construction, $U$ and $V$ are symmetric, e.g. $U_{jk} =U_{kj} $.}
Now, projecting equation \eqref{eigenval-s} on $\langle\emptyset|$, we get
\beq
\mu(s) = (e^{-s}-1)(T_{i+1}-U_{i,i+1})+(e^{s}-1)(T_{i}-U_{i,i+1}).
\label{eq:mus}
\eeq
We also project equation \eqref{eigenval-s} on $\langle\{j\} |$ for $j=1$, $1<j<i$ and $i+1<j<L$, $j=i$, $j=i+1$,  and $j=L$. We get respectively:
\bea
\mu(s)\,T_1 &=&\alpha (1-T_1)-\gamma T_1 +\lambda^2(1-T_1-T_2)+T_2-T_1
\nonu 
&&+ (e^{-s}-1)(U_{1,i+1}-V_{1,i,i+1})
+(e^{s}-1)(U_{1,i}-V_{1,i,i+1})\,,
\label{eq:T1s}\\
\mu(s)\,T_j &=& \lambda^2(2-2T_j-T_{j+1}-T_{j-1}) +T_{j+1}-2T_j+T_{j-1} 
\nonu
&&+ (e^{-s}-1)(U_{j,i+1}-V_{j,i,i+1})+(e^{s}-1)(U_{j,i}-V_{j,i,i+1})\,,
\label{eq:Tjs}\\
\mu(s)\,T_i &=& \lambda^2(2-2T_i-T_{i+1}-T_{i-1}) +T_{i+1}-2T_i+T_{i-1} 
\nonu
&&+ (e^{-s}-1)(T_{i+1}-U_{i,i+1})\,,
\label{eq:Tis}\\
\mu(s)\,T_{i+1} &=& \lambda^2(2-2T_{i+1}-T_{i+2}-T_{i}) +T_{i+2}-2T_{i+1}+T_{i} 
\nonu
&&+(e^{s}-1)(T_{i}-U_{i,i+1})\,,
\label{eq:TIs}\\
\mu(s)\,T_L &=&\delta (1-T_L)-\beta T_L +\lambda^2(1-T_L-T_{L-1})+T_{L-1}-T_L
\nonu 
&&+ (e^{-s}-1)(U_{i+1,L}-V_{i,i+1,L})
+(e^{s}-1)(U_{i,L}-V_{i,i+1,L})\,.
\label{eq:TLs}
\eea
These equations are solved iteratively, expanding all quantities as series in $s$.
We set 
\begin{eqnarray*}
 \mu(s) & = & \mu^{(0)}+s\,\mu^{(1)}+\frac{s^2}{2}\mu^{(2)}+o(s^2)\,,\\
 T_j(s) & = & T_j^{(0)}+s\,T_j^{(1)}+o(s)\,, \\
 U_{j,k}(s) & = & U_{j,k}^{(0)}+s\,U_{j,k}^{(1)}+o(s)\,.
\end{eqnarray*}
In the above expansions, $\mu^{(0)}=0$ is the greatest eigenvalue of the undeformed Markov matrix and $\mu^{(1)}=\langle J^{lat}_{i\rightarrow i+1} \rangle$ is the mean value of the lattice current measured between the site $i$ and $i+1$, where the deformation occurs.
We recall that $\langle J^{lat}_{i\rightarrow i+1} \rangle$ has been computed in \eqref{mean:lattice}. The value of 
$T_j^{(0)}= \langle n_j \rangle$ has also been already calculated, see \eqref{1pt-corr-bis}. Similarly, $U_{j,k}^{(0)}$ is linked to the two-points correlation function, see \eqref{2pt-corr-bis}.

We wish to compute $\mu(s)$ up to order 2, which corresponds to the variance of the lattice current. We get it through the expansion of \eqref{eq:mus} up to order 2:
\bea
\mu^{(1)} &=& T_i^{(0)}-T_{i+1}^{(0)}\,,\label{eq:mu1}
\\
\mu^{(2)} &=& 2\Big( T_i^{(1)}-T_{i+1}^{(1)} \Big) +T_i^{(0)}+T_{i+1}^{(0)} -2\,U_{i,i+1}^{(0)}\,.
\label{eq:mu2}
\eea
Equation \eqref{eq:mu1} just reproduces the relation \eqref{eq:drho-discret} between the mean values of the lattice current and of the density. 

To get $T_j^{(1)}$, one considers equations \eqref{eq:T1s}-\eqref{eq:TLs} at first order in $s$. They only involve  
$T_j^{(1)}$, $T_j^{(0)}$ and $U_{jk}^{(0)}$, and can be solved recursively in $T_j^{(1)}$. We get
\begin{equation*}
\left(\begin{array}{c} T_{i+1}^{(1)} \\[1.2ex] T_{i}^{(1)} \end{array}\right)=\frac{\phi^L}{1-ab\phi^{2L-2}}
\left(\begin{array}{cc} \displaystyle
\frac{b\phi^{L-i-1}+\phi^{i+1-L}}{\phi^2-1}&\displaystyle \frac{a\phi^{i}+\phi^{-i}}{\phi^2-1}
\\[2.1ex]
\displaystyle\frac{ b\phi^{L-i}+\phi^{i-L}}{\phi^2-1}&\displaystyle \frac{a\phi^{i-1}+\phi^{1-i}}{\phi^2-1}\end{array}\right)
 \,\left(\begin{array}{c} \sum_{l=0}^{i-1}\Big(a\phi^l+\phi^{-l}\Big)I_{l+1} \\[2ex] \sum_{l=0}^{L-i-1}\Big(b\phi^l+\phi^{-l}\Big)I_{L-l} \end{array}\right) 
\end{equation*}
with
\bea
I_{j} &=&\frac{(\phi+1)^2}{4\phi}\times \begin{cases}
\mu^{(1)}\, T_i^{(0)}+T_{i+1}^{(0)}-U_{i,i+1}^{(0)}\,,\quad &\mbox{for } j=i
\\[1ex]
\mu^{(1)}\, T_{i+1}^{(0)}-T_{i}^{(0)}+U_{i,i+1}^{(0)}\,,\quad &\mbox{for } j=i+1
\\[1ex]
\mu^{(1)}\, T_j^{(0)}+U_{j,i+1}^{(0)}-U_{j,i}^{(0)}\,,\quad &\mbox{otherwise}
\end{cases}
\eea
Plugging these values into \eqref{eq:mu2}, we get the analytical expression of the variance of the lattice current:
\bea
&&\mu^{(2)}\ =\  T_i^{(0)}+T_{i+1}^{(0)}-2\,U_{i,i+1}^{(0)}\label{mu2} \\
&&+ \frac2{1+\phi}\,\left\{ \frac{\phi^i(b\phi^{2L-2i-1}-1)}{1-ab\phi^{2L-2}}\, \sum_{\ell=0}^{i-1} (a\phi^\ell+\phi^{-\ell})I_{\ell+1}
-\frac{\phi^{L-i}(a\phi^{2i-1}-1)}{1-ab\phi^{2L-2}}\, \sum_{\ell=0}^{L-i-1} (b\phi^\ell+\phi^{-\ell})I_{L-\ell}\right\}.
\nonumber
\eea
Using the explicit form of $I_j$, one can compute the sums in \eqref{mu2} to perform the thermodynamic limit for $\mu^{(2)}$, see section \ref{sect:thermo}.
Let us conclude this subsection by mentioning that the higher cumulants may be computed in principle by similar methods. However, the computations become much harder and are 
beyond the scope of this paper.

\subsection{Comparison with SSEP \label{comparison-SSEP}}

As mentioned previously the {DiSSEP} is a deformation of the SSEP which can be easily recovered when taking $\lambda=0$.
This limit already reveals at the level of the matrix product ansatz algebra: the commutation relations between $E$, $D$ and $H$ \eqref{eq:comEDH}
become simpler when $\lambda=0$. We have indeed $[H,E]=[H,D]=0$. Hence $H$ can be chosen equal to the identity. In this case we recover the well known
relation $[D,E]=D+E$, relevant in the construction of the steady state of the SSEP. Remark that the generators $G_1$, $G_2$ and $G_3$ in contrast
``diverge'' when taking the limit $\lambda \rightarrow 0$ (this can be seen by inverting the change of basis \eqref{change-basis}).

We can also take the limit $\lambda \rightarrow 0$ at the level of the physical observables.
For the one and two points correlation function we get
 \begin{equation*}
 \lim\limits_{\lambda \rightarrow 0} \ \langle n_i \rangle = \frac{\rho_a\left(L+B-i\right)+\rho_b\left(i-1+A\right)}{L+A+B-1},
 \mb{with} A=\frac{1}{\alpha+\gamma}\,,\ B=\frac{1}{\beta+\delta}
 \end{equation*}
and 
\begin{equation*}
 \lim\limits_{\lambda \rightarrow 0} \ \langle n_i n_j \rangle_c = -\frac{\left(i-1+A\right)\left(B+L-j\right)(\rho_a-\rho_b)^2}
 {\left(L+A+B-1\right)^2\left(L+A+B-2\right)},
\end{equation*}
which are in agreement with the known expressions for the SSEP \cite{DDR,DerrReview}.
For the lattice and evaporation-condensation currents we get
\begin{equation*}
 \lim\limits_{\lambda \rightarrow 0} \ \langle J^{lat}_{i\rightarrow i+1} \rangle = \frac{\rho_a-\rho_b}{L+A+B-1}
 \quad \text{and} \quad \lim\limits_{\lambda \rightarrow 0} \ \langle J^{cond}_{i, i+1} \rangle = 0,
\end{equation*}
also in agreement with the SSEP results.

We can also take carefully the limit of the expression \eqref{mu2} to recover the variance of the lattice current for the SSEP model \cite{DDR}
\begin{align}
&\mu_{2,SSEP} = \frac{\rho_a+\rho_b}{L+A+B-1}+\frac{(A-3A^2+2A^3+B-3B^2+2B^3)(\rho_a-\rho_b)^2}{3(L+A+B-1)^3(L+A+B-2)}
\nonu
&\!\!
-\frac{(\rho_a-\rho_b)^2}{3(L+A+B-1)^2(L+A+B-2)}+\frac{\rho_a^2+\rho_b^2}{(L+A+B-1)(L+A+B-2)}-\frac{2(\rho_a^2+\rho_a\rho_b+\rho_b^2)}{3(L+A+B-2)}.\quad
\nonumber
\end{align}

\section{Thermodynamic limit\label{sect:thermo}}

In this section, we study the thermodynamic limit of the previous model when there exists a competition between the diffusion of particles and 
the evaporation/condensation of pairs.

\subsection{Scaling of the parameters\label{sect:scaling}}

In order to maintain the competition in the continuous limit, we have to scale properly the parameters of the model.
In other words, the mean time for a particle to go through the lattice by diffusion must be comparable to the time 
for it to be evaporated.

Let us write the time evolution of the one point correlation function for $1<i<L$
\begin{eqnarray}
 \frac{d \langle n_i \rangle}{dt} & = & \langle n_{i-1}(1-n_i)\rangle + \langle (1-n_i)n_{i+1} \rangle -\langle (1-n_{i-1})n_i\rangle - \langle n_i(1-n_{i+1})\rangle 
 \nonu
 & & +\lambda^2\Big( \langle (1-n_{i-1})(1-n_i)\rangle + \langle (1-n_i)(1-n_{i+1})\rangle - \langle n_{i-1}n_i\rangle -  \langle n_in_{i+1}\rangle \Big)\label{eq:den1}\\
 & = & \langle n_{i-1} \rangle + \langle n_{i+1} \rangle -2\langle n_i \rangle
 +\lambda^2\Big(2-\langle n_{i-1} \rangle - \langle n_{i+1} \rangle -2 \langle n_i \rangle \Big).\label{eq:den}
\end{eqnarray}
Note that although the two-point correlation functions cancel when going from \eqref{eq:den1} to \eqref{eq:den},   
the mean field approximation is not exact in the sense that the connected two-point function does not vanish, see \eqref{2pt-corr-bis}.

\null

 \textsl{Remark:} The one point correlation function verifies a closed set of equation as for the SSEP, in contrast with the ASEP case 
where the equations couple the one point function and the two points function. This property remains valid for the higher order
correlation functions \eqref{eq:reccurence-corr}, which allows in principle to compute them. However, for the multi-points correlation functions, 
solving this set of equation can be very hard. This points out the usefulness of the matrix product ansatz which makes the calculations much easier.

\null

We want to take the large $L$ limit in equation \eqref{eq:den}. We set $i=Lx$ with $x\in [0\,,\,1]$ and $\langle n_i \rangle = \rho(x)$.
We get
\begin{equation}
 \frac{d\rho}{dt}(x)=\frac{1}{L^2}(\rho''(x)+o(1))+2\lambda^2(1-2\rho(x)+o(1))
\end{equation}
We see on the previous equation that we have to take $\lambda=\lambda_0/L$ in order to have a balance between diffusion and creation-annihilation.
After a rescaling of the time $t \rightarrow L^2 t$, we obtain
\begin{equation} \label{eq:MF}
 \frac{\partial \rho}{\partial t}= \frac{\partial^2 \rho}{\partial x^2}+2\lambda_0^2(1-2\rho)
\end{equation}
with the boundary conditions $\rho(0)=\rho_a$ and $\rho(1)=\rho_b$. This equation shows that the correlation length for this scaling
is finite. Indeed, the stationary density obtained in equation \ref{1pt-corr-limit} decays as
$\overline{\rho}(x)-1/2\sim \exp(-2\lambda_0x)$ for $x$ far from the boundaries. The correlation length can thus be defined as $1/(2\lambda_0)$.

Without rescaling $\lambda$ and the time w.r.t $L$ (or for $\lambda=\lambda_0/L^\mu$ with $\mu<1$ and rescaling the time $t \rightarrow L^{2\mu} t$ ), 
the diffusive term drops out and the density satisfies
\begin{equation} \label{eq:MF-1}
 \frac{\partial \rho}{\partial t}= 2\lambda^2(1-2\rho)\;.
\end{equation}
In the case where $\lambda=\lambda_0/L^\mu$ for  $\mu>1$ with the rescaling of the time $t \rightarrow L^{2} t$, 
the system becomes a pure diffusive model for large $L$  and one gets for the density
\begin{equation} \label{eq:MF-2}
 \frac{\partial \rho}{\partial t}= \frac{\partial^2 \rho}{\partial x^2}\;.
\end{equation}

\subsection{Thermodynamic limit of the observables}

From \eqref{1pt-corr}, we can compute the expression of the one point correlation function in the continuous limit ($i=Lx$ and $\lambda=\lambda_0/L$)
\begin{equation} \label{1pt-corr-limit}
\overline{\rho}(x):= \lim\limits_{L\rightarrow \infty} \langle n_{Lx} \rangle 
=\frac{1}{2}+\frac{1}{2\sinh 2\lambda_0} \left(q_1 e^{-2\lambda_0(x-1/2)}+q_2 e^{2\lambda_0(x-1/2)} \right)\;,
\end{equation}
where
\begin{equation}\label{def:q1}
q_1=\left(\rho_a+\rho_b-1\right)\sinh(\lambda_0)-\left(\rho_b-\rho_a\right)\cosh(\lambda_0),
\end{equation}
and 
\begin{equation}\label{def:q2}
q_2=\left(\rho_a+\rho_b-1\right)\sinh(\lambda_0)+\left(\rho_b-\rho_a\right)\cosh(\lambda_0)\;.
\end{equation}
It is easy to check that it satisfies the stationary version of \eqref{eq:MF}. 

We can also compute the two point correlation function in this limit. One can see that it scales as $\frac1L$, i.e. it has weak correlations. We get:
\begin{eqnarray}
& & c_2(x,y) = \lim\limits_{L\rightarrow \infty} L \times \langle n_{Lx}n_{Ly} \rangle_c =  
\frac{2\lambda_0  \,q_1q_2}{\left(\sinh 2\lambda_0 \right)^3}\,\sinh 2\lambda_0(1-y) \sinh 2\lambda_0 x \,.\qquad
\end{eqnarray}
For $\lambda_0<<1$, this two-point correlation function behaves algebraically w.r.t. $x$ and $y$ whereas it behaves exponentially and 
is short range for $\lambda_0>>1$.

The limit of the particle currents are given by
\begin{equation}
 j^{lat}(x):= \lim\limits_{L\rightarrow \infty} L \times \langle J^{lat}_{Lx\rightarrow Lx+1} \rangle =
 \frac{\lambda_0}{\sinh 2\lambda_0} \left(q_1 e^{-2\lambda_0(x-1/2)}-q_2 e^{2\lambda_0(x-1/2)} \right)\;,
\end{equation}
and 
\begin{equation}
 j^{cond}(x):= \lim\limits_{L\rightarrow \infty} L^2 \times \langle J^{cond}_{Lx, Lx+1} \rangle =
  \frac{-\lambda_0^2}{\sinh 2\lambda_0} \left(q_1 e^{-2\lambda_0(x-1/2)}+q_2 e^{2\lambda_0(x-1/2)} \right)\;.
\end{equation}
Remark that these expressions are consistent with the fact that when the system reaches a thermodynamic equilibrium,
that is for $\rho_a=\rho_b=1/2$ (or equivalently $q_1=q_2=0$), both currents vanish.

The particle conservation law \eqref{eq:part.conservation-discret} becomes in the thermodynamic limit
\begin{equation*}
 -\frac{dj^{lat}}{dx}(x)+2j^{cond}(x)=0,
\end{equation*}
which is satisfied by the expressions above.
In the same way, relations \eqref{eq:drho-discret} and \eqref{jcond-discret} become in the thermodynamic limit:
\bea\label{relation:rho-cond}
\frac{d\overline\rho}{dx}(x)+j^{lat}(x)=0\,,\qquad j^{cond}(x)=\lambda_0^2\Big(1-2\overline\rho(x)\Big).
\eea

\subsubsection*{Behavior of the density and the currents}
\begin{figure}[pht]
  \begin{center}
    \subfloat[Density]{
      \includegraphics[width=0.4\textwidth,height=0.3\textheight]{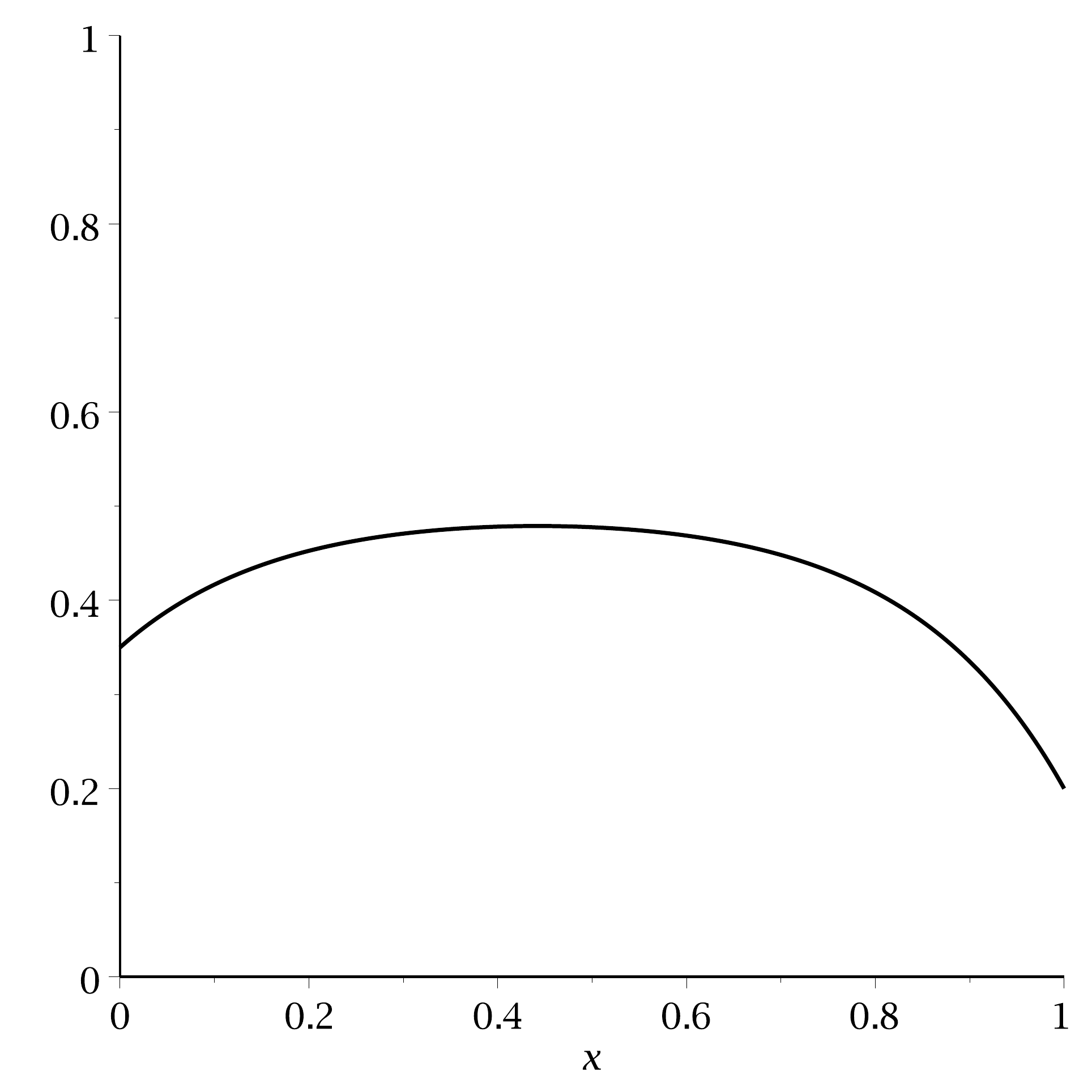}
      \label{sub:densite_concave}
                         }\hfill
    \subfloat[Mean value (---) and variance ($\cdots$) of the lattice current]{
      \includegraphics[width=0.4\textwidth,height=0.3\textheight]{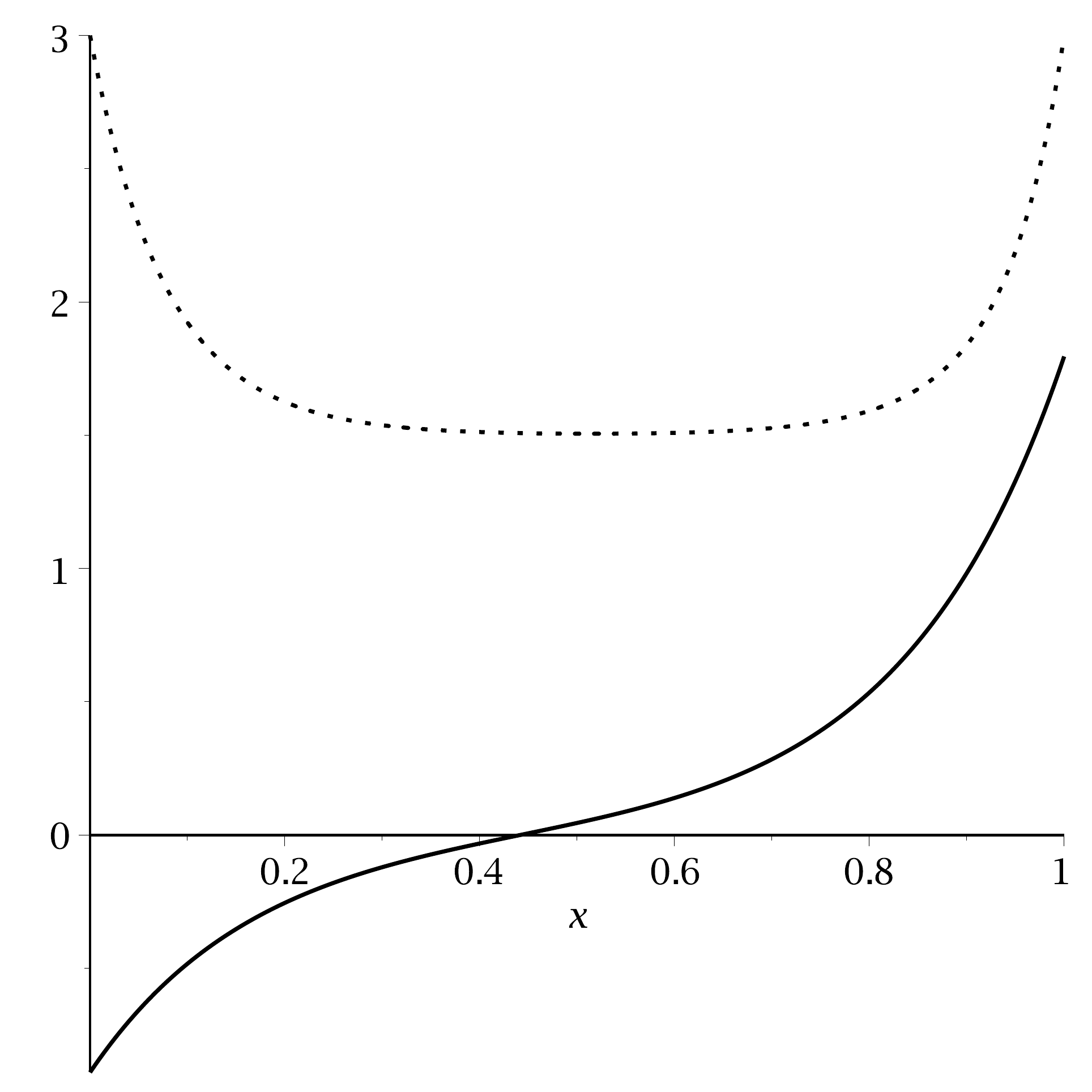}
      \label{sub:courant_concave}
                         }
    \caption{Plot of the density and of the lattice current for $\rho_a=0.35$, $\rho_b=0.2$ and $\lambda_0=3$.}
    \label{fig:concave}
  \end{center}
\end{figure}

\begin{figure}[pht]
  \begin{center}
    \subfloat[Density]{
      \includegraphics[width=0.4\textwidth,height=0.3\textheight]{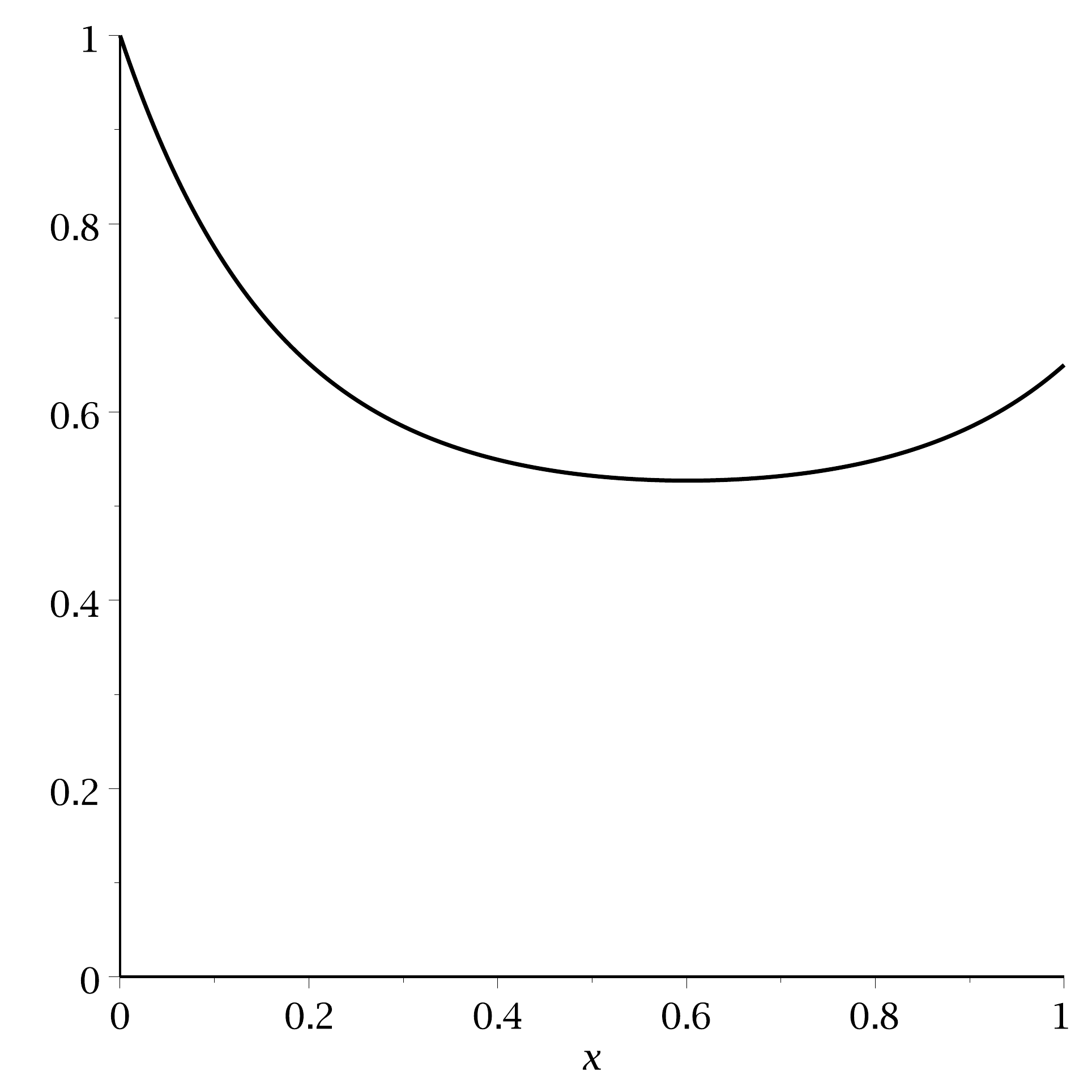}
      \label{sub:densite_convexe}
                         }\hfill
    \subfloat[Mean value (---) and variance ($\cdots$) of the lattice current]{
      \includegraphics[width=0.4\textwidth,height=0.3\textheight]{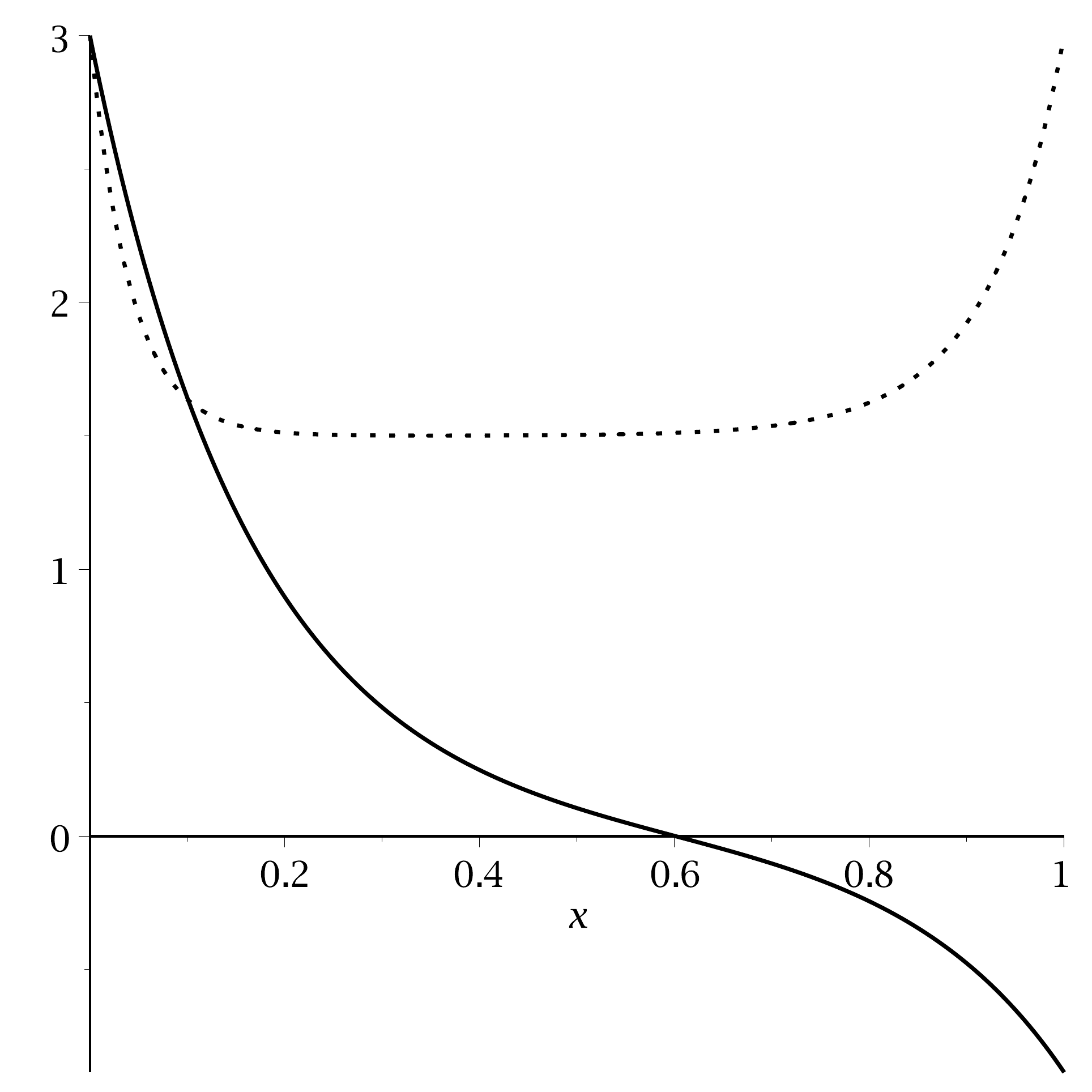}
      \label{sub:courant_convexe}
                         }
    \caption{Plot of the density and of the lattice current for $\rho_a=1$, $\rho_b=0.65$ and $\lambda_0=3$.}
    \label{fig:convexe}
  \end{center}
\end{figure}

\begin{figure}[pht]
  \begin{center}
    \subfloat[Density]{
      \includegraphics[width=0.4\textwidth,height=0.3\textheight]{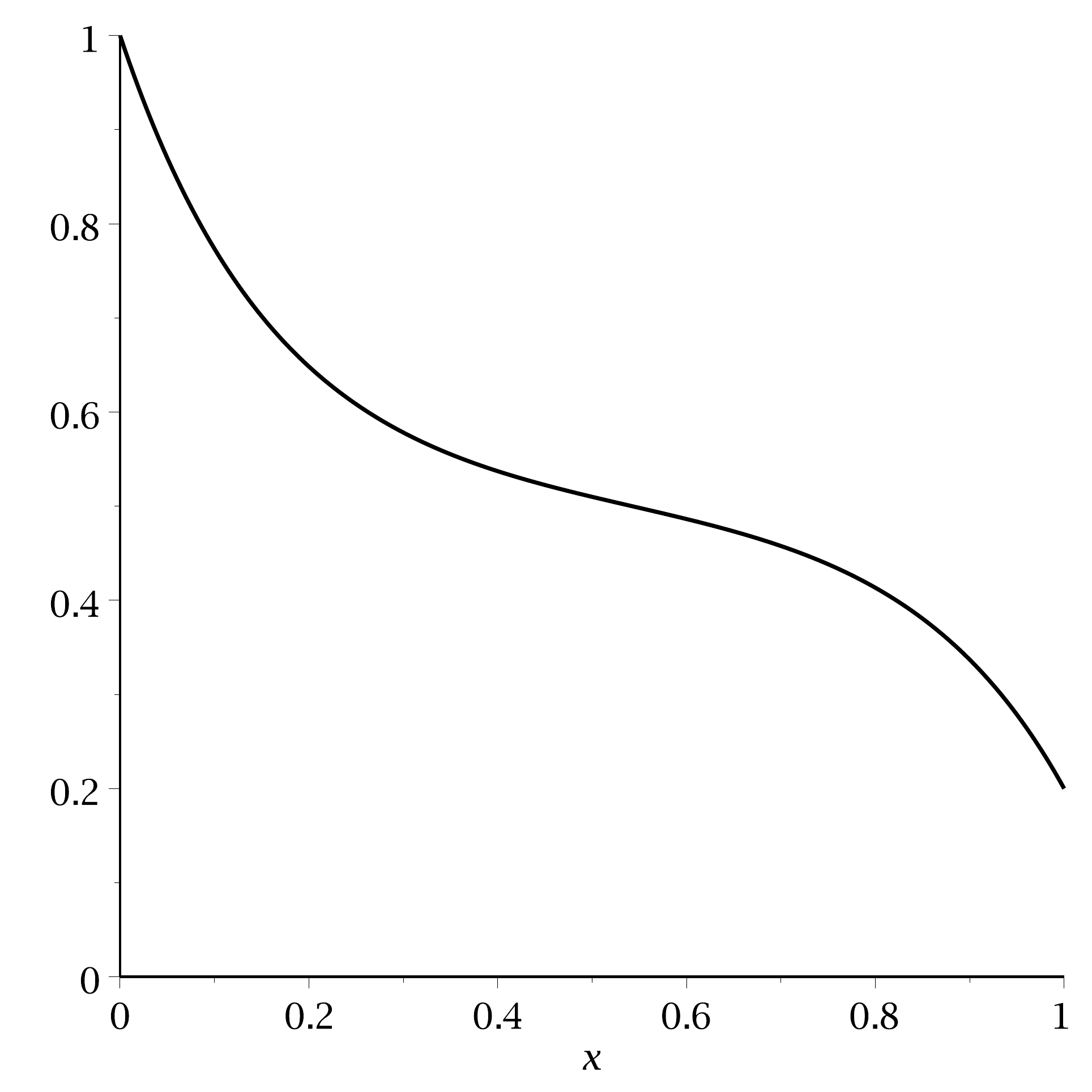}
      \label{sub:densite_monotone}
                         }\hfill
    \subfloat[Mean value (---) and variance ($\cdots$) of the lattice current]{
      \includegraphics[width=0.4\textwidth,height=0.3\textheight]{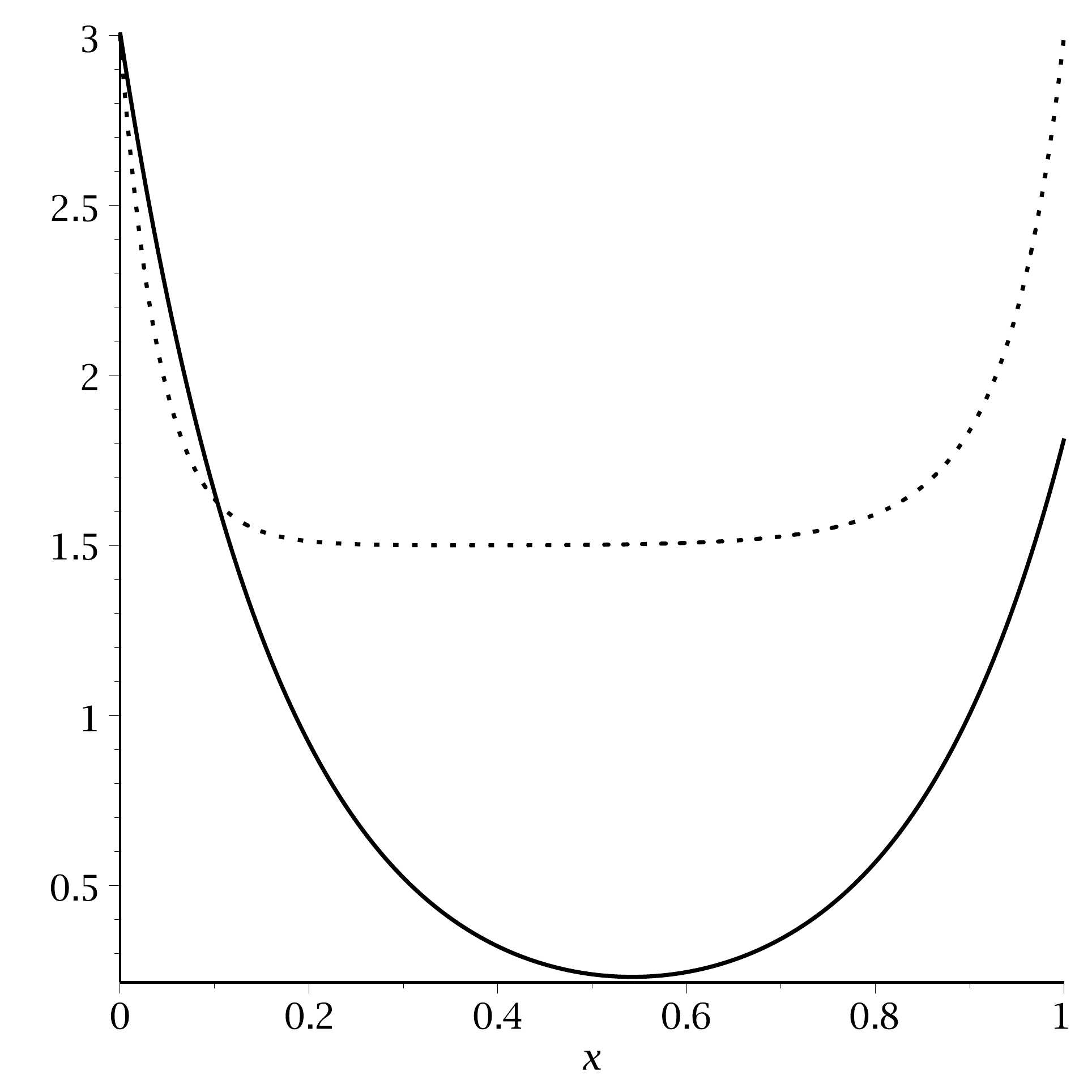}
      \label{sub:courant_monotone}
                         }
    \caption{Plot of the density and of the lattice current for $\rho_a=1$, $\rho_b=0.2$ and $\lambda_0=3$.}
    \label{fig:monotone}
  \end{center}
\end{figure}

Depending on the values of $q_1$ and $q_2$ defined in \eqref{def:q1} and \eqref{def:q2}, the behavior of the density may change: 
 \begin{itemize}
 \item the density is not monotonic when $e^{-2\lambda_0}< \frac{q_1}{q_2}< e^{2\lambda_0}$, which implies that $q_1$ and $q_2$ have the same sign. In that case, it possesses an extremum
 at $\overline{x}$ 
 satisfying $e^{4\lambda_0(\overline x -1/2)}=\frac{q_1}{q_2}$. The lattice current vanishes at this point.  
 \begin{itemize}
 \item the density presents a maximum
 \begin{equation}
  \overline{\rho}(\overline x)=\frac12-\frac{\sqrt{q_1q_2}}{\sinh(2\lambda_0)}
 \end{equation}
 when $q_1,q_2<0$. Let us remark that in this case, the density is everywhere smaller than $1/2$.
 Example of such behavior can be seen on figure \ref{fig:concave}.
 
The lattice current changes direction at the point $\overline x$, as expected since the lattice current 
 goes from high density to low density.
 At this point, the condensation current is minimal but positive, since the density is smaller than $1/2$, so that condensation is promoted.
 \item It presents a minimum
 \begin{equation}
  \overline{\rho}(\overline x)=\frac12+\frac{\sqrt{q_1q_2}}{\sinh(2\lambda_0)}
 \end{equation}
 when $q_1,q_2>0$. In this case, the density is everywhere greater than $1/2$. 
 
 The condensation current is negative but maximal, so that the evaporation is minimal. As previously, the lattice current changes sign at $\overline x$, still going from high density to low density. 
  Example of such behavior can be seen on figure \ref{fig:convexe}.
 \end{itemize}
 \item The density is monotonic from $\rho_a$ to $\rho_b$ when $\frac{q_1}{q_2}<e^{-2\lambda_0} $ or $\frac{q_1}{q_2} > e^{2\lambda_0}$.
 In this case, the lattice current never vanishes.  Example of such behavior can be seen on figure \ref{fig:monotone}.

\end{itemize}
The condensation current follows the same pattern, due to the relation \eqref{relation:rho-cond}.
The lattice current behaves as follows:
 \begin{itemize}
 \item it is not monotonic when $e^{-2\lambda_0}< -\frac{q_1}{q_2}< e^{2\lambda_0}$, which implies that $q_1$ and $q_2$ have opposite sign. 
There is an extremum at $\overline{x}$ 
 satisfying $e^{4\lambda_0(\overline x -1/2)}=-\frac{q_1}{q_2}$. The condensation current vanishes at this point.  
 \begin{itemize}
 \item  When $q_1<0$, the lattice current presents a maximum
 \begin{equation}
  j^{lat}(\overline x)=-\frac{2\lambda_0}{\sinh(2\lambda_0)}\sqrt{-q_1q_2}.
 \end{equation} 
 \item When $q_1>0$, it presents a minimum (see figure \ref{fig:monotone})
 \begin{equation}
 j^{lat}(\overline x)=\frac{2\lambda_0}{\sinh(2\lambda_0)}\sqrt{-q_1q_2}.
 \end{equation}
 \end{itemize}
 \item The lattice current is monotonic  when $-\frac{q_1}{q_2}<e^{-2\lambda_0} $ or $-\frac{q_1}{q_2} > e^{2\lambda_0}$, see figures \ref{fig:concave} and \ref{fig:convexe}.
\end{itemize}

\subsubsection*{Variance of the lattice current}
The thermodynamic limit of the variance of the lattice current, computed exactly in \eqref{mu2} for any size, takes the form:
\bea\label{eq:mumicro}
\mu_2(x) &=& 2q_1\,q_2\,\lambda_0^{2}\left\{(2x-1)\,\frac {\sinh\big( 2\,\lambda_0\,(2\,x-1)\big) }{ \left( \sinh( 2\,\lambda_0)  \right) ^{3}}
- \frac { \cosh( 2\,\lambda_0) \,\cosh\big( 2\,\lambda_0\,(2x-1)\big)
 +1}{ \left( \sinh( 2\,{\lambda_0})  \right) ^{4}}\right\}
 \nonu
&-&\!\!
q_2^{2}\lambda_0\,{\frac{{e^{4\,\lambda_0\,x}}+
{e^{-4\,\lambda_0\, ( 1-x ) }} -{e^{4\,\lambda_0\, \left( 2\,x-1 \right) }}+3}{4 \left( \sinh( 2\,\lambda_0) 
 \right) ^{3}}}
 -q_1^{2}\lambda_0\,{\frac{e^{4\,\lambda_0\, ( 1-x ) }+e^{-4\,\lambda_0\,x}-e^{4\,\lambda_0\, ( 1-2\,x ) } +3
 }{4 \left( \sinh( 2\,\lambda_0)  \right) ^{3}}}
 \nonu
&+&\frac{\lambda_0\,\cosh
 \left( 2\,\lambda_0\,x \right) \cosh \big( 2\,\lambda_0\,( 1-x )  \big) }{\sinh( 2\,\lambda_0) }.
\eea
As all physical quantities of the model, the variance is invariant under the transformation $q_1\,\leftrightarrow\,q_2$ and $x\to 1-x$, which is the left-right symmetry. The particle-hole symmetry amounts to change $q_1\to-q_1$ and $q_2\to-q_2$: it leaves $\mu_2$ invariant, transforms $\overline\rho(x)$ into $1-\overline\rho(x)$ and changes the sign of the currents. The symmetry $\lambda\to-\lambda$ reads $\lambda_0\to-\lambda_0$ and $q_1\leftrightarrow(-q_2)$ and leaves all quantities invariant.

\subsubsection*{SSEP limit}
Finally, by taking the limit $\lambda_0 \rightarrow 0$ in the previous quantities, we recover the well-known SSEP expressions \cite{DDR}:
\begin{eqnarray}
&&\overline\rho_{{SSEP}}(x)= \rho_a(1-x)+\rho_b x,\qquad
c_{2,SSEP}(x,y) =  -x(1-y)(\rho_a-\rho_b)^2,\\
&& j^{lat}_{{SSEP}}(x) = \rho_a-\rho_b,\qquad\qquad\quad\
\mu_{2,SSEP}(x) = \rho_a+\rho_b -\frac23(\rho_a^2+\rho_a\rho_b+\rho_b^2).
\end{eqnarray}

\section{Comparison with Macroscopic fluctuation theory\label{sect:MFT}}

\subsection{Presentation of the Macroscopic fluctuation theory}
The model presented in this paper belongs to a larger class of models describing lattice gas with diffusive dynamics and
evaporation-condensation in the bulk which are driven out of equilibrium by two reservoirs at different densities. 
In the thermodynamic limit, the dynamics of these models can be understood (under proper assumptions) through the so-called
Macroscopic Fluctuation Theory (MFT). More precisely the large deviation functional for the density profile and the particle currents
has been computed in \cite{BL,BL2} based on the pioneering works for diffusive models \cite{Bertini1,Bertini2,GianniRevue}. 
The aim of this section is to extract from this general framework the local variance of the current on
the lattice for the DiSSEP model and check its exact agreement with the value of $\mu_2(x)$ computed previously, see eq. \eqref{eq:mumicro} from the microscopic
point of view.

Let us start by briefly presenting the key ingredients of the MFT related to our model. A detailed presentation can be found in \cite{BL,BL2}.
It has been shown that the microscopic behavior of the system can be averaged in the thermodynamic limit and can be described at the 
macroscopic level by a small number of relevant parameters: $D(\rho)$, $\sigma(\rho)$, $A(\rho)$ and $C(\rho)$.
These parameters depend on the microscopic dynamics of the model and have to be computed for each different model. 
The two first are related to the diffusive dynamics on the lattice: $D(\rho)$ is the diffusion coefficient and $\sigma(\rho)$ is
the conductivity. For the DiSSEP, the diffusive dynamics is the same as for the SSEP and hence these coefficients take the 
values $D(\rho)=1$ and $\sigma(\rho)=2\rho(1-\rho)$.
The two other parameters $A(\rho)$ and $C(\rho)$ are related to the creation-annihilation dynamics. $A(\rho)$ can be understood 
intuitively as the mean number of particles annihilated per site and per unit of time when the density profile is identically flat
and equal to $\rho$ in the system whereas $C(\rho)$ stands for the mean number of particles created.
A rigorous definition of these parameters can be found in \cite{BL}. 
For the DiSSEP, we have $A(\rho)=2\lambda_0^2\rho^2$ and $C(\rho)=2\lambda_0^2(1-\rho)^2$.
When the number of sites $L$ goes to infinity, the probability of observing a given history of the density profile $\rho$, of the lattice current
$\dot{Q}$ and of the creation-annihilation current $\dot{K}$ during the time interval $[0,T]$\footnote{We keep here all the notations
used in \cite{BL}. The link with the quantities previously computed is given by the fact that in the stationary state the mean value of
$\dot{Q}(x)$ is $j^{lat}(x)$ and the mean value of $\dot{K}(x)$ is $2j^{cond}(x)$.}, can by written as
\bea \label{eq:large_devia}
 \mathbb{P}_{[0,T]}\left( \{ \rho,\dot{Q},\dot{K} \} \right) \sim \exp \left[-L\mathcal{I}_{[0,T]}(\rho,\dot{Q},\dot{K}) \right],
\eea
with the large deviation functional
\bea
 \mathcal{I}_{[0,T]}(\rho,\dot{Q},\dot{K})=\int_0^T dt \int_0^1 dx \left\{\frac{(\dot{Q}(x,t)+D(\rho(x,t))\partial_x\rho(x,t))^2}{2\sigma(\rho(x,t))}
 + \Phi\left(\rho(x,t),\dot{K}(x,t)\right)\right\},
\eea
where
\bea\label{eq:Phi}
 \Phi(\rho,\dot{K})=\frac{1}{2}\left[ A(\rho)+C(\rho)-\sqrt{\dot{K}^2+4A(\rho)C(\rho)}+\dot{K}\ln\left(\frac{\sqrt{\dot{K}^2+4A(\rho)C(\rho)}+\dot{K}}{2C(\rho)}\right)\right].
\eea
The factor $1/2$ in the definition of $\Phi$ is a slight modification in comparison to \cite{BL} due to the fact that we consider here
creation-annihilation of pairs of particles instead of creation-annihilation of single particles.

The quantities $\rho$, $\dot{Q}$ and $\dot{K}$ are related through the conservation equation
\bea
\partial_t \rho(x,t)=-\partial_x \dot{Q}(x,t)+\dot{K}(x,t),
\eea
and the value of $\rho$ is fixed on the boundaries $\rho(0,t)=\rho_a$, $\rho(1,t)=\rho_b$.
The minimum of the large deviation functional $\mathcal{I}_{[0,T]}$ is achieved when the particle currents take their typical values, that is
 $\dot{Q}(x,t)=-D(\rho(x,t))\partial_x\rho(x,t)$ and $\dot{K}(x,t)=C(\rho(x,t))-A(\rho(x,t))$. The typical evolution of the density profile is
 hence given by
 \bea
 \partial_t \rho(x,t)=\partial_x \left( D(\rho(x,t))\partial_x\rho(x,t)\right)+C(\rho(x,t))-A(\rho(x,t))
 \eea
 which matches \eqref{eq:MF} for the DiSSEP.

\subsection{Computation of the variance of the lattice current}

Using the previous formalism and following \cite{BL}, it is possible to compute the local variance of the lattice current $\dot{Q}$ in the stationary regime. 
Due to the fact that the dynamics of the model does not conserve the number of particle, this current and its fluctuations depend on the
position in the system. Hence, given a function $\tau(x)$, we want to compute the generating function of the cumulants of the integrated current
$\int_0^T dt \int_0^1 dx\; \tau(x)\; \dot{Q}(x,t)$ for $T$ going to infinity.
\bea
\mathcal{F}(\{ \tau \} ) = \lim\limits_{T \rightarrow \infty}\lim\limits_{L \rightarrow \infty} \frac{1}{LT}
\ln \mathbb{E}_{[0,T]}\left( \exp \left( \int_0^T dt \int_0^1 dx \tau(x)\dot{Q}(x,t) \right) \right).
\eea 
The previous expression can be simplified using \eqref{eq:large_devia} and a saddle point method. It reduces to maximize a functional
over the time dependent fields $\rho$, $\dot{Q}$ and $\dot{K}$. Assuming that the extrema of this functional is achieved for time independent
profiles, we end up with the following expression (the reader is invited to refer to \cite{BL} for the details)
\bea
\mathcal{F}(\{ \tau \} ) = \sup\limits_{\rho,\dot{Q}} \left( \int_0^1 dx \tau(x)\dot{Q}(x)-\hat{\mathcal{I}}(\rho,\dot{Q}) \right),
\eea 
with
\bea
\hat{\mathcal{I}}(\rho,\dot{Q})= \int_0^1 dx \left( \frac{(\dot{Q}(x)+D(\rho(x))\partial_x\rho(x))^2}{2\sigma(\rho(x))}
 + \Phi\left(\rho(x),\partial_x\dot{Q}(x)\right) \right).
\eea 
To compute the local variance of the lattice current $\dot{Q}$ at the point $y$, it is enough to take $\tau(x)=\delta(x-y)$ and expand 
$\mathcal{F}(\{ \epsilon \tau \} )$ up to order $\epsilon^2$. For a small perturbation $\epsilon$, the fields are expected to be close to
their typical value
\bea
\rho(x)=\overline{\rho}(x)+\epsilon\frac{f(x)}{D(\overline{\rho}(x))}, \qquad \dot{Q}(x)=-D(\overline{\rho}(x))\partial_x \overline{\rho}(x)+\epsilon q(x)
\eea 
with the constraint $f(0)=f(1)=0$ due to the boundaries.
We then obtain
\bea\label{eq:la}
\mathcal{F}(\{ \epsilon \tau \} )  =  -\epsilon D(\overline{\rho}(y))\partial_x \overline{\rho}(y) + \frac{\epsilon^2}{2}\mu_2(y) 
\eea
with the variance of the lattice current at the point $y$ 
\bea
 \mu_2(y) =2\sup\limits_{q,f} \left\{ q(y)-\int_0^1 dx\left( \frac{(q(x)+f'(x))^2}{2\sigma(\overline{\rho}(x))}+\frac{(q'(x)+U(x)f(x))^2}{4(A(\overline{\rho}(x))+C(\overline{\rho}(x)))} \right) \right\}
\eea
and $U(x)=\frac{A'(\overline{\rho}(x))-C'(\overline{\rho}(x))}{D(\overline{\rho}(x))}=4\lambda_0^2$\;.
We make the following change of variables to solve this optimization problem
\bea\label{eq:defpsi}
\varphi(x)=\frac{q(x)+f'(x)}{\sigma(\overline{\rho}(x))}, \qquad \psi(x)=\frac{q'(x)+U(x)f(x)}{2(A(\overline{\rho}(x))+C(\overline{\rho}(x)))},
\eea
so that the Euler-Lagrange equations become for the DiSSEP:
\begin{equation}
 \left\{ \begin{aligned}
         & \psi'(x)=\varphi(x)-\delta(x-y) \\
         & \varphi'(x)=4\lambda_0^2 \psi(x).
         \end{aligned} \right. 
\end{equation}
Note that there are slight modifications in expressions \eqref{eq:la} and \eqref{eq:defpsi} with respect to \cite{BL}, in accordance with the modification of $\Phi$ 
(see discussion after \eqref{eq:Phi}).

These equations can be solved analytically and we get
\begin{equation}
 \left\{ \begin{aligned}
         & \psi(x)=\frac{\theta(x\leq y)\sinh(2\lambda_0 x)\cosh(2\lambda_0(1-y))+\theta(x>y)\sinh(2\lambda_0(x-1))\cosh(2\lambda_0 y)}{\sinh 2\lambda_0}\\
         & \varphi(x)=\frac{2\lambda_0\left[\theta(x\leq y)\cosh(2\lambda_0 x)\cosh(2\lambda_0(1-y))+\theta(x>y)\cosh(2\lambda_0(x-1))\cosh(2\lambda_0 y) \right]}{\sinh 2\lambda_0}.
         \end{aligned} \right. 
         \label{eq:phipsi}
\end{equation}
The function $q(x)$ can be also computed analytically by solving 
\begin{equation}
q''(x)-4\lambda_0^2 q(x)= \partial_x(2(A(\overline{\rho}(x))+C(\overline{\rho}(x)))\psi(x))-4\lambda_0^2 \sigma(\overline{\rho}(x))\varphi(x)\;.
\end{equation}
Note that it depends on $y$, see for instance the expressions \eqref{eq:phipsi}. It allows us to deduce the expression of $q(x)$ at the special point $y$ (as needed in \eqref{eq:la})
\begin{equation}
 q(y)= \int_0^1 dx \left[ \sigma(\overline{\rho}(x))\varphi(x)^2+2(A(\overline{\rho}(x))+C(\overline{\rho}(x)))\psi(x)^2  \right], 
\end{equation}
with $\varphi$ and $\psi$ are given above. Hence for the DiSSEP, the variance of the current lattice computed from MFT is 
\begin{eqnarray*}
 \mu_2(y)
 & = & \int_0^1 dx \left[ \sigma(\overline{\rho}(x))\varphi(x)^2+2(A(\overline{\rho}(x))+C(\overline{\rho}(x)))\psi(x)^2  \right] \\
 & = & \frac{4\lambda_0^2}{\sinh^2 2\lambda_0}\left[ \cosh^2 2\lambda_0(1-y) \int_0^1 dx \left( \sigma(\overline{\rho}(x))+\sinh^2 2\lambda_0x \right) \right. \\
 & & \qquad \qquad \qquad \left.+\cosh^2 2\lambda_0y \int_0^1 dx \left( \sigma(\overline{\rho}(x))+\sinh^2 2\lambda_0(1-x) \right) \right]\;.
\end{eqnarray*}
Using the explicit form for $\sigma$, we show that this result obtained from MFT matches perfectly the previous result \eqref{eq:mumicro} computed exactly 
from a microscopic description of the model.

Let us remark that a similar comparison between MFT for diffusive model \cite{Bertini1,Bertini2,GianniRevue} and microscopic exact computations 
has been performed in \cite{DCairns} for the SSEP.
The result obtained here is a confirmation of the MFT developed in \cite{BL,BL2} for a system with diffusion and dissipation.

\section{Spectrum of the Markov matrix \label{sec:bethe}}

\subsection{Link with the XXZ spin chain, integrability and Bethe equations}

The model introduced above (DiSSEP) possesses the distinctive feature of being integrable, i.e. the Markov matrix $M$ governing the process belongs
 to a set of commuting operators. Let us recall briefly, the main objects to get this set. The detailed construction for this particular 
 model can be found in \cite{CRV}. This set is constructed through a generating operator depending on a spectral parameter,
 the so-called transfer matrix $t(x)$. The building blocks of this transfer matrix are the $R$-matrix, obeying the Yang-Baxter equation, and the 
 boundary matrices $K$ and $\widetilde K$ satisfying respectively the reflection equation and the dual reflection equation. These equations ensures 
 the commutation of the transfer matrix for different value of the spectral parameter as it was shown in \cite{sklyanin}: $[t(x),t(y)]=0$.
 The Markov matrix is then obtained as the first moment of the transfer matrix: $M\propto t'(1)$.
 
The integrability of this model is also revealed through its unexpected connexion with the XXZ model. To be more precise, let us introduce the 
following Hamiltonian $H$
\begin{eqnarray}
  H&=&(\alpha-\gamma)\sigma_1^+ -\frac{\alpha+\gamma}{2}(\sigma^z_1+1)\ +\ (\delta-\beta)\sigma_L^+ -\frac{\delta+\beta}{2}(\sigma^z_L+1)\nonumber\\
  &&-\frac{\lambda^2-1}{2}\ \sum_{k=1}^{L-1}
 \Big( \sigma_k^x\sigma_{k+1}^x+\sigma_k^y\sigma_{k+1}^y-\frac{\lambda^2+1}{\lambda^2-1}(\sigma_k^z\sigma_{k+1}^z-1) \Big)\label{eq:Hxxz}
\end{eqnarray}
where $\sigma^{x,y,z,+,-}$ are the Pauli matrices. It corresponds to the open XXZ spin chain with upper triangular boundaries.
This Hamiltonian $H$ is conjugated to the Markov matrix $M$ defined in \eqref{Markov-matrix}. Namely, one has
\begin{equation}\label{eq:sim}
  H=Q_1Q_2\dots Q_L M Q_1^{-1}Q_2^{-1}\dots Q_L^{-1}\quad\text{where}\quad Q=\begin{pmatrix} -1&1\\1&1 \end{pmatrix} \;.
\end{equation}
Let us also mention that the XXZ Model for particular choices of boundaries is conjugated to the Markov matrix of the open ASEP. However, for the boundaries 
present in \eqref{eq:Hxxz}, the conjugation provides non-Markovian boundaries.  

We deduce from \eqref{eq:sim} that the spectrum of $M$ is identical to the one of $H$. Moreover, 
the eigenvalues (but not the eigenvectors) of XXZ spin chain with upper triangular boundaries are the same that the ones for diagonal boundaries and one 
can use the results of \cite{Gaudin,ABBBQ,sklyanin}. 
Let us mention that equality between the spectrums of two different models has been used previously to study models with only evaporation \cite{Alcaraz1}.
Note also that for $\lambda^2=1$, the bulk Hamiltonian becomes diagonal, and the full Hamiltonian triangular, allowing to get its spectrum easily without Bethe ansatz, in accordance with the results of section \ref{sect:lambda1}.

The eigenvalues of $H$ with diagonal boundaries
can be parametrized in two different ways depending on the choice of the pseudo-vacuum:
\begin{itemize}
 \item For the pseudo-vacuum with all the spins up and in the notations of the present paper,
 the eigenvalues of $H$ are given by
 \begin{equation}\label{eq:ei1}
  E=-\alpha-\beta-\gamma-\delta+4(\phi-1)^2 \sum_{i=1}^N \frac{u_i}{(u_i-\phi^2)(u_i-1)}
 \end{equation}
where $N=0,1,\dots,L$ and $u_i$ are the Bethe roots. The Bethe roots must satisfy the following Bethe equations
\begin{eqnarray}\label{eq:be1}
\frac{u_i+a\phi^2}{\phi(au_i+1)}\ \frac{u_i+b\phi^2}{\phi(bu_i+1)} \left(\frac{\phi(u_i-1)}{u_i-\phi^2} \right)^{2L}=
\prod_{j=1 \atop j\neq i}^N \frac{\phi^2(\phi^2u_i-u_j)(u_iu_j-1)}{(u_i-\phi^2u_j)(u_iu_j-\phi^4)}
\end{eqnarray}
where $i=1,2,\dots,N$ and $a$ and $b$ are defined in \eqref{eq:bordsG}.
\item
For the pseudo-vacuum with all the spins down, the eigenvalues of $H$ are given by
 \begin{equation}\label{eq:ei2}
  E=4(\phi-1)^2 \sum_{i=1}^N \frac{v_i}{(v_i-\phi^2)(v_i-1)}
 \end{equation}
where $v_i$ satisfy the following Bethe equations
\begin{eqnarray}\label{eq:be2}
\frac{av_i+\phi^2}{\phi(v_i+a)}\ \frac{bv_i+\phi^2}{\phi(v_i+b)} \left(\frac{\phi(v_i-1)}{v_i-\phi^2} \right)^{2L}=
\prod_{j=1 \atop j\neq i}^N \frac{\phi^2(\phi^2v_i-v_j)(v_iv_j-1)}{(v_i-\phi^2v_j)(v_iv_j-\phi^4)}\;.
\end{eqnarray}
\end{itemize}

Let us stress again that, although the spectrum of the XXZ spin chain is the same for diagonal or upper boundaries, the eigenvectors are different. 
For the XXZ spin chains with upper triangular boundaries, the eigenvectors associated to the parametrization \eqref{eq:ei1} and \eqref{eq:be1} of the eigenvalues 
were computed only recently by algebraic Bethe ansatz in \cite{PL,Bel} based on the previous results for the XXX spin chain \cite{CR,BCR,BC}.
The computation of the eigenvectors associated to the parametrization \eqref{eq:ei2} and \eqref{eq:be2} is still an open problem.

\subsection{Computation of the spectral gap \label{sec:gap}}

In this section, we want to study the dynamical properties of the model: using the previous Bethe equations, 
we study the approach to the stationary state at large times for a large system. We must compute the eigenvalue, denoted by $G$, 
for the first excited state (\textit{i.e.} the one with the 
greatest non-vanishing eigenvalue). 

\begin{figure}[htb]
\begin{center}
 \begin{tikzpicture}[scale=5]
\draw [->] (-1,0) -- (1,0)  ;
\draw [->] (0,-1) -- (0,1)  ;
\node [below right] at (1,0) {$a$};
\node [above left] at (0,1) {$b$};
 \node at (-0.05,-0.05) {0};
\draw[thick, domain=0.0625:1,samples=300] plot (\x, { 0.25^2/\x   });
\draw[thick] (-1,0.25) -- (0.25,0.25) -- (0.25,-1) ;
\node at (0.6,-0.6) {$G=\frac{-4(\phi-1)^2a}{(\phi^2+a)(a+1)}$};
\node at (0.6,0.6) {$G=-2\frac{\phi-1}{\phi+1}\left(\frac{a-1}{a+1}+\frac{b-1}{b+1}\right)$};
\node at (-0.6,0.6) {$G=\frac{-4(\phi-1)^2b}{(\phi^2+b)(b+1)}$};
\node at (-0.5,-0.5) {$G=-4\frac{(|\phi|-1)^2}{(\phi+1)^2}$};
\draw[dotted] (0.25,0) --(0.25,0.25) --(0,0.25)  ;
\node [below] at (0.3,0) {$|\phi|$};
\node [left] at (0,0.3) {$|\phi|$};
 \end{tikzpicture}
 \end{center}
\caption{Value of the gap G depending on the parameters $a$, $b$ and $\phi$. The equation of the curve is $b=\phi^2/a$.
(This particular figure is drawed for $\phi=1/4$ ($\lambda=0.6$) even though similar one is valid for any $\phi$).}
 \label{fig:g3}
\end{figure}
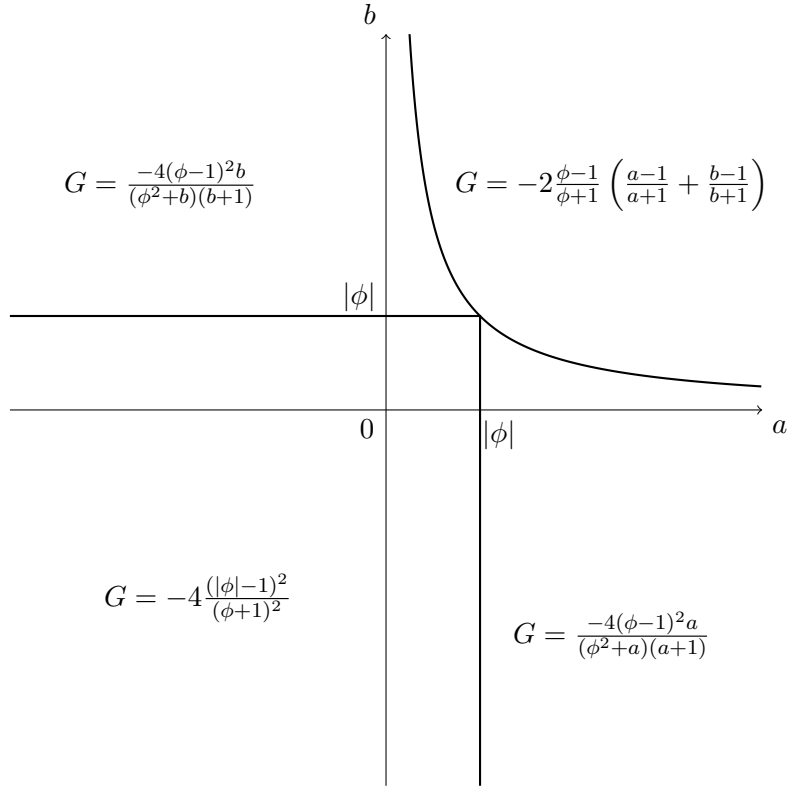

\begin{figure}[htb]
\begin{center}
\begin{minipage}{0.6\textwidth}
 \includegraphics[width=\textwidth,height=0.4\textheight]{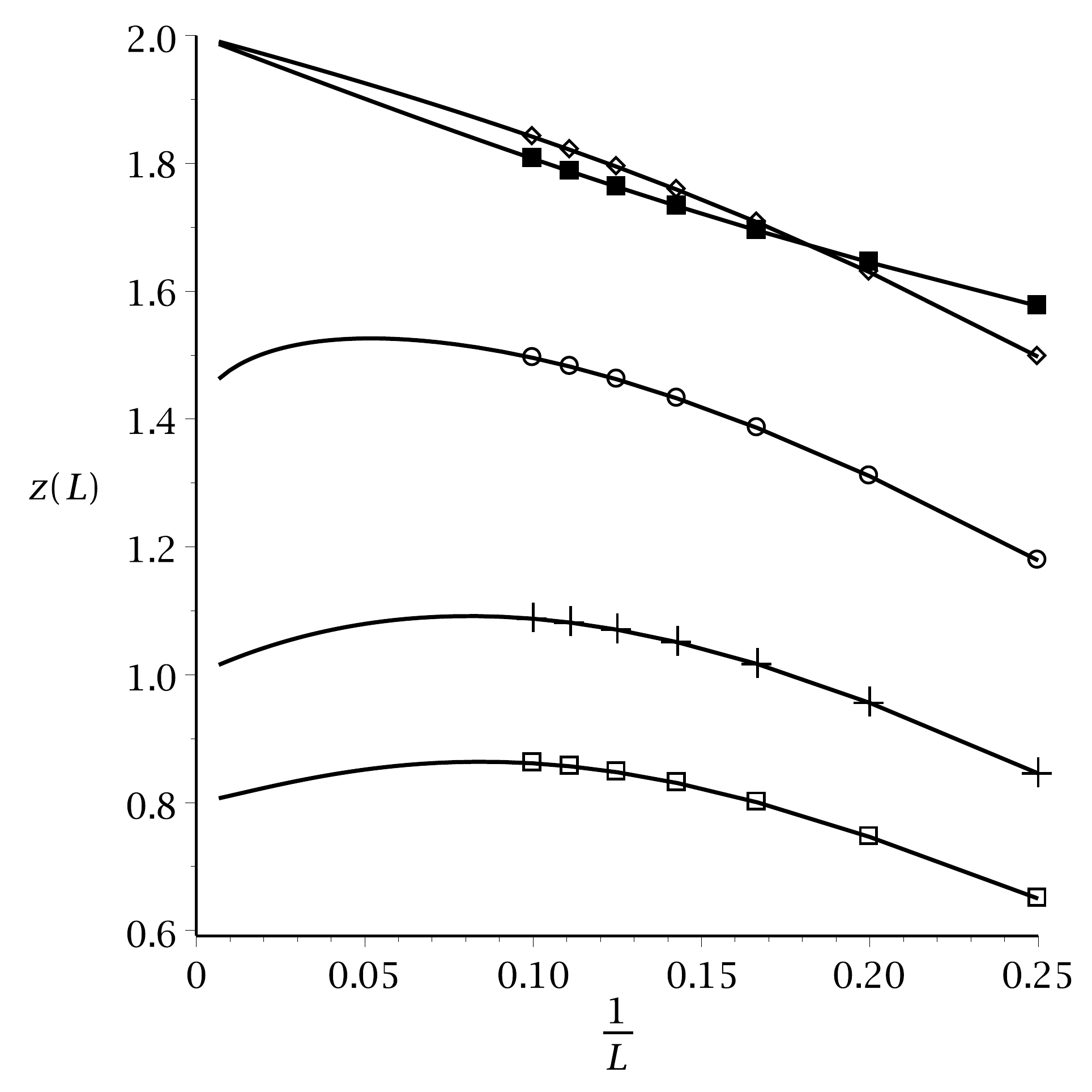}
\end{minipage}
\begin{minipage}{0.2\textwidth}
 \begin{tikzpicture}[scale=2]
\node at (.2,.7) {$\rule{2mm}{2mm}: \ \mu=2.$};
\node at (.2,.2) {$\Diamond :\ \mu=1.$};
\node at (.2,-.5) {$\circ :\ \mu=0.7$};
\node at (.2,-1.2) {$+: \ \mu=0.5$};
\node at (.2,-1.7) {$\Box: \ \mu=0.4$};
 \end{tikzpicture}
\end{minipage}
\end{center}
\caption{Behavior of the gap in the thermodynamic limit when $\lambda$ behaves as $\frac{1}{L^{\mu}}$: plot 
of $z(L)=\frac{\ln(G_L)-\ln(G_{L-1})}{\ln(L-1)-\ln({L})}$ as a function of $\frac1L$.
The lines represent the values obtained from Bethe ansatz (for $4\leq L\leq  150$), while the dots correspond to direct diagonalisation of $H$
(for $4\leq L\leq10$).}
 \label{fig:mu}
\end{figure}

We start by presenting the main results for the gap then we give the sketch of the numerical evidences for them.
\begin{itemize}
\item In the case when evaporation rate $\lambda$ is independent of the size of the system $L$, there is a non-vanishing gap. The values of this gap depends on
the boundaries parameters and on $\lambda$. We present these different values of the gaps on Figure \ref{fig:g3}. 
They are consistent with the analytical result obtained for $\lambda=1$ (see section \ref{sect:lambda1}).
\item
If the rate $\lambda$ behaves as $\frac{1}{L^\mu}$ for large system, the model is gapless and we get
\begin{equation}\label{eq:gap}
 G\sim \frac{1}{L^{2\mu}} \mb{for} 0<\mu<1 \mb{and}  G\sim \frac{1}{L^{2}} \mb{for} \mu\geq 1\;.
\end{equation}
We show in figure \ref{fig:mu} numerical evidence for such a behavior. We plot $z(L)=\frac{\ln(G_L)-\ln(G_{L-1})}{\ln(L-1)-\ln({L})}$ as a function of $\frac1L$: as $\frac1L$ tends to 0, it tends to $2\mu$ (resp. 2) for $\mu<1$ (resp. $\mu\geq1$).
The $\frac{1}{L^{2}}$ behavior of the gap for $\mu> 1$ is expected since 
the system becomes in this case a diffusive model in the thermodynamic limit as discussed in section \ref{sect:scaling}.
\end{itemize}

To prove these result, we must study in detail the Bethe equations \eqref{eq:be1} and \eqref{eq:be2}.
The comparison of the eigenvalues obtained by the exact diagonalisation of M or by the numerical resolutions of the Bethe equations for small system (up to $10$ sites),
 show that the gap is obtained for $N=1$ in \eqref{eq:ei2} and \eqref{eq:be2} or is equal to $G=-\alpha-\beta-\gamma-\delta=-2\frac{\phi-1}{\phi+1}\left(\frac{a-1}{a+1}+\frac{b-1}{b+1}\right)$ 
(which corresponds to $N=0$ in \eqref{eq:ei1} and \eqref{eq:be1}).
We assume that this behavior holds for any $L$ then we must solve only \eqref{eq:be1} for $N=1$.
This Bethe equation can be written as the vanishing of a polynomial of degree $2L+2$ w.r.t. $v_1$.
This polynomial has two obvious roots $\phi$ and $-\phi$ which are not physical since they corresponds to a vanishing ``eigenvector''. The remaining factor
is a polynomial of degree $2L$ w.r.t. $v_1$ which can be transformed, thanks to \eqref{eq:ei2} (and up to a normalization), to a polynomial of degree $L$ w.r.t. $E$.
Then, the Bethe equation \eqref{eq:be2} for $N=1$ becomes
\begin{eqnarray}\label{eq:GE}
 \sum_{p=0}^L \frac{(1+\phi)^{2p}\ E^p}{4^p}&&\sum_{q=0}^{L-p}\phi^{2q}\left[
 ab\ \begin{pmatrix}p+q\\q\end{pmatrix}\begin{pmatrix} L-q-2\\p-2\end{pmatrix}+
 (a+b)\ \begin{pmatrix}p+q-1\\q\end{pmatrix}\begin{pmatrix} L-q-1\\p-1\end{pmatrix}\right. \nonumber\\
 &&\left.\qquad\qquad +\begin{pmatrix}p+q-2\\q\end{pmatrix}\begin{pmatrix} L-q\\p\end{pmatrix}
 \right]=0
\end{eqnarray}
The L.H.S. of the previous equation is a factor of the characteristic polynomial of the Hamiltonian $H$ \eqref{eq:Hxxz} or of the Markov matrix $M$ \eqref{Markov-matrix}.
It is possible now to find numerically the roots of the polynomial \eqref{eq:GE} for large system (up to $150$ sites) and pick up the largest ones.
Performing this computation for different values of $\lambda$ and of the boundary parameters, we obtain the results for the gap summarized previously, see figure \ref{fig:g3}.  

\section{Conclusion\label{conclu}}
The DiSSEP model (see Fig. \ref{fig:rdm}) was recently introduced in \cite{CRV}. It was 
shown that the model is integrable and that the probability distribution 
function describing the stationary state can be written using the matrix 
product Ansatz. In the present paper we discuss in detail the properties 
of the model. If one chooses the symmetric hopping rate equal to 1, the 
physics is dependent on the parameter $\lambda$ whose square is the common rate
for annihilation and creation of pair of particles.

 It can be shown the the Hamiltonian (Markov matrix) has the same spectrum 
as a XXZ spin 1/2 quantum chain with non-diagonal boundary terms (see 
Eq. \eqref{eq:Hxxz}). One observes that if $\lambda$ vanishes, one gets the ferromagnetic
XXX model which corresponds to the well known SSEP model. A natural idea is to
study the system in the weak dissipation limit $\lambda\, \sim\, 1/L^\mu$ where $L$ 
is the size of the system. We have studied the effect of this Ansatz on the 
physics of the model. We remind the reader that the weak ASEP model \cite{DEL} is 
defined in a similar way. The size dependence enters in the forward-backward
hopping asymmetry

 As a warm-up exercise we have studied in detail the $\lambda = 1$ case. In 
the bulk, the Hamiltonian is diagonal in this case and the spectrum can be 
easily be computed. The current large deviation function on the first bond 
has been derived. The function is convex and behaves like $j\ln(j)$ for large 
values of the current $j$.

 The case where $\lambda$ is arbitrary was considered next. Using the 
matrix product Ansatz, we have obtained the expression \eqref{eq:G123} for the 
average value of an arbitrary monomial of the generators of the 
quadratic algebra \eqref{eq:comG} in a specific representation \eqref{eq:bordsG}. 
This expression allows to compute any correlator. 
We give the expressions for the average values of the local density 
\eqref{1pt-corr}, of the two-point and three-point densities correlators \eqref{2pt-corr-bis} 
respectively \eqref{3pt-corr}. We also show (see Eq. \eqref{eq:dens}) that to derive the 
average density profile, the mean-field result is exact. There are two 
kinds of currents in our problem. The first one is given by particles crossing
a bond between two sites, the second one is given by the pair of particles 
which leave or enter the system by the  creation and annihilation 
processes. Both expressions are given (see Eqs. \eqref{mean:lattice} and \eqref{mean:cond}). The
variance of the first current was also computed, the result can be found in
Eq. \eqref{mu2}.   

 In order to study the properties of the model in the large $L$ limit, as 
mentioned before, we take $\lambda$ vanishing like $L^{-\mu}$. Using Eqs. 
\eqref{1pt-corr} and \eqref{2pt-corr-bis} one sees that the correlation length is proportional to 
1/$\lambda$ which suggests that the system is gapless for $\mu > 0$. Using 
the Bethe Ansatz, we have shown that this is indeed the case. The energy 
gap behavior is given in Eq. \eqref{eq:gap}. We could be tempted to look closely at 
the value $\mu$ = 1/2 when the gap vanishes like $1/L$, suggesting conformal 
invariance. This is not the case as one can see from the behavior of the 
average density and the two-point correlation function which do not have the 
expected behavior \cite{AR}. We have decided to consider in detail the case 
$\mu$ = 1 which corresponds to a critical dynamic exponent $z = 2$ 
corresponding to diffusive processes. The results are given in Section \ref{sect:thermo}.

We have also compared our results with 
those which can be obtained using the Macroscopic Fluctuation Theory 
\cite{BL,BL2}. The variance of the current computed using this method coincides 
with the lattice calculation described earlier in the text.

Finally, we would like to point out two generalizations  which look to us interesting. The first one is to consider asymmetric hopping rates. The system will probably be not integrable but mean-field and Monte Carlo simulations will reveal new physics. The second generalization deals with the multi-species problem keeping integrability. This implies a generalization of the results obtained in \cite{CRV}.

\subsection*{Acknowledgements:} It is a pleasure to warmly thank L. Ciandrini, C. Finn for fruitful discussions and suggestions.

\end{document}